\documentclass[multphys,vecphys]{svmult}

\usepackage{makeidx}         
\usepackage{graphicx}        
\usepackage{multicol}        
\usepackage[bottom]{footmisc}
\usepackage{amssymb,amsfonts}
\usepackage{mathrsfs}
\usepackage{wasysym}

\makeindex             

\newcommand{\bit}{\begin{itemize}}
\newcommand{\eit}{\end{itemize}}
\newcommand{\be}{\begin{equation}}
\newcommand{\ee}{\end{equation}}
\newcommand{\bea}{\begin{eqnarray}}
\newcommand{\eea}{\end{eqnarray}}
\newcommand{\nn}{\nonumber}

\newcommand{\tgm}{\tilde\gm}

\newcommand{\der}[2]{\frac{\partial #1}{\partial #2}}
\newcommand{\dert}[2]{{\partial #1 / \partial #2}}

\newcommand{\Sp}{{\mathcal S}}

\newcommand{\Df}{\mathcal{D}}

\newcommand{\gam}{\gamma}
\newcommand{\gm}{\gamma}

\begin{document}

\title*{Mass and Angular Momentum in General Relativity}
\author{Jos\'e Luis Jaramillo\inst{1,2} \and Eric Gourgoulhon\inst{2}}

\institute{
Instituto de Astrof\'\i sica de Andaluc\'\i a,
CSIC, Apartado Postal 3004, Granada 18080, Spain 
\texttt{jarama@iaa.es}
\and 
Laboratoire Univers et Th\'eories, Observatoire de
  Paris, CNRS, Universit\'e Paris Diderot, 5 place Jules Janssen,
  F-92190 Meudon, France
\texttt{eric.gourgoulhon@obspm.fr}
}
%
%
\maketitle

We present an introduction to mass and angular
momentum in General Relativity. After briefly reviewing energy-momentum
for matter fields, first in the flat Minkowski case (Special Relativity) 
and then in curved spacetimes with or without symmetries, we 
focus on the discussion of energy-momentum for the gravitational
field. We illustrate the difficulties rooted in the Equivalence Principle
for defining a local energy-momentum density for the gravitational
field. This leads to the understanding of  gravitational energy-momentum
and angular momentum as non-local observables that make sense, at best,
for extended domains of spacetime. After introducing Komar quantities 
associated with spacetime symmetries, it is shown how
total energy-momentum can be  unambiguously defined for isolated systems,  
providing fundamental tests for the internal consistency of General 
Relativity as well as setting the conceptual basis for the understanding
of energy loss by gravitational radiation.
Finally, several attempts to formulate quasi-local notions
of mass and angular momentum associated with extended but finite
spacetime domains are presented, together with some illustrations
of the relations between total and
quasi-local quantities in the particular context of black hole
spacetimes. This article is not intended to be a rigorous and exhaustive
review of the subject, but rather an invitation to the topic
for non-experts. In  this sense we follow essentially
the expositions in \cite{Szaba04,Gou07a,Poiss04,Wald84} and  
refer the reader interested in further developments to the existing literature,
in particular to the excellent and comprehensive review by Szabados \cite{Szaba04}.

\section{Issues around the notion of gravitational energy
in General Relativity}

\subsection{Energy-momentum density for matter fields}
Let us first consider mass and angular momentum associated with matter in the 
absence of gravity, in a flat Minkowski spacetime.
The density of energy and linear momentum associated with a distribution of matter
are encoded in the energy-momentum tensor $T_{\mu\nu}$, corresponding to 
the Noether current conserved under infinitesimal spacetime 
translations in a Lagrangian framework.
This general conservation property,  namely $\partial_\mu T^{\mu\nu}=0$  
in inertial Minkowski coordinates,
plays a key role in our discussion. Indeed,  together with the presence of symmetries, it
permits the introduction of conserved quantities or {\em charges}.
Given a space-like hypersurface $\Sigma$ and considering the unit time-like vector $n^\mu$ normal to 
it, we can define the conserved quantity 
associated with the symmetry $k^\mu$ and the domain $D$ ($\subset \Sigma$) as
\be
\label{e:conservedcharges}
Q_D[k^\mu]=\int_D k^\rho T_{\nu\rho} n^\nu \sqrt{\gamma} \; d^3 x \ \ ,
\ee 
where  $\sqrt{\gamma}\; d^3x$ denotes the induced volume element in $D$. 
The conservation of $T_{\mu\nu}$ and the characterisation of
$k^\mu$ as a symmetry imply the conservation of the vector ${T^\mu}_\nu k^\nu$,
i.e. $\partial_\mu \left({T^\mu}_\nu k^\nu\right)=0$. Applying then the Stokes theorem, it follows
 the equality between the
change in time of $Q_D[k^\mu]$ and the flux of ${\gamma^\mu}_\rho T^{\rho\nu}k_\nu$ through
the boundary of $D$ (where ${\gamma^\mu}_\nu$ is the projector on $D$).
Minkowski spacetime symmetries are
given by Poincar\'e transformations. Therefore, we can associate conserved quantities
with the infinitesimal generators corresponding to translations  $\mathrm{T}^\nu_{a}$,
rotations $\mathrm{J}^\mu_i$, and boosts $\mathrm{K}^\mu_i$ (here the label $a$ for translation 
generators runs in $\{0,1,2,3\}$, whereas  
$i$ is a space-like index in $\{1,2,3\}$). In this manner, a 4-momentum
$P_a[D]$ and an angular momentum $J_i[D]$ associated with the  
distribution of matter in $D\subset \Sigma$ can be defined as
\be
\label{e:matterdensities1}
P_a[D]=\int_D T_{\mu\nu} \mathrm{T}^\nu_{a} n^\mu \sqrt{\gamma} \; d^3 x \ \ , \ \
J_{i}[D]=\int_D T_{\mu\nu} \mathrm{J}_{i}^\nu n^\mu \sqrt{\gamma} \; d^3 x \ \ .
\ee
More generally, we can combine together the rotation and boost generators  
$\mathrm{J}^\mu_i$ and $\mathrm{K}^\mu_i$ into a 
vector-field-valued antisymmetric matrix $\mathrm{M}_{[ab]}^\mu$
(where $\mathrm{J}^\mu_i = {{}^3\!\epsilon_i}^{jk}\mathrm{M}_{[jk]}^\mu$
and $\mathrm{K}^\mu_i=\mathrm{M}_{[0i]}^\mu$) and write the conserved
quantities
\be
\label{e:matterdensities2}
J_{[ab]}[D]=\int_D T_{\mu\nu} \mathrm{M}_{[ab]}^\nu n^\mu \sqrt{\gamma} \; d^3 x \ \ .
\ee
The mass and (Pauli-Lubanski) spin are constructed as
\bea
\label{e:CasimirPoicare}
m^2[D] := -\eta^{ab} P_a[D]P_b[D] \ \ , \ \ S^a[D]:= \frac{1}{2} 
{}^4\!\epsilon^{abcd} P_b[D] J_{[cd]}[D] \ \ ,
\eea
in terms of which Poincar\'e Casimirs (invariant under Poincar\'e transformations)
can be expressed.

In the non-flat case,  (matter) energy-momentum 
tensor acts as the source of gravity through the Einstein equation and,
consistently with Bianchi identities, satisfies the divergence-free
condition analogous to the flat conservation law:
\be
\label{e:Einstein}
G_{\mu\nu}:={}^4\!R_{\mu\nu} -\frac{1}{2}{}^4\!R \; g_{\mu\nu}= 8\pi T_{\mu\nu} \ \ , \ \ 
\nabla_\nu T^{\nu\mu} = 0 \ º .
\ee
The same strategy employed in the flat case for defining physical quantities associated with matter,
i.e. from conserved currents corresponding to some symmetry, can be followed in 
non-flat spacetimes $({\cal M}, g_{\mu\nu})$ presenting  Killing vectors $k^\mu$.
The vector ${T^\mu}_\nu k^\nu$ is conserved, 
i.e. $\nabla_\mu \left( {T^\mu}_\nu k^\nu \right)=0$, and
provides a current-density for the conserved quantity
$Q_D[k^\mu]$ defined by expression (\ref{e:conservedcharges}).
The physical interpretation of $Q_D[k^\mu]$ depends of course on the nature of the Killing 
vector $k^\mu$.
Actually, $Q_D[k^\mu]$ does not actually depend on the slice $\Sigma$ in the 
sense that its value is the same in the domain of dependence of $D$ (this precisely 
corresponds to the conserved
nature of this charge).

In a general spacetime with no symmetries the previous strategy ceases to work,
and ambiguities in the definition of mass and angular momentum enter into scene.
One can still calculate the flux of ${T^\mu}_\nu \xi^\nu$
for a given vector $\xi^\nu$, and define the associated quantity $Q_D[\xi^\mu]$.
However, the latter will now depend on the slice $\Sigma$ and, in addition, its explicit 
dependence on $\xi^\mu$ introduces some degree of arbitrariness in the discussion.
In this context, given a space-like 3+1 foliation $\{\Sigma_t\}$ of the
spacetime with time-like normal vector $n^\mu$, the current $P^\mu:= -T^{\mu\nu}n_\nu$
can be interpreted as the energy-momentum density associated
with (Eulerian) observers {\em at rest} with respect to $\Sigma_t$. That is, 
$E:= T^{\mu\nu}n_\mu n_\nu$ stands as the matter energy density and 
$p^\mu:=- {\gamma^\mu}_\rho T^{\rho\nu}n_\nu$
as the momentum density, where $\gamma_{\mu\nu}$ is the induced metric on $\Sigma_t$
(see Eq. (\ref{e3+1Tmunu}) below for the complete 3+1 decomposition of $T_{\mu\nu}$).
In particular, we can calculate the matter energy associated with
observers $n^\mu$ over the spatial region $D$ by direct integration
\be 
\label{e:matterenergy}
E[D]= \int_D E \sqrt{\gamma}\; d^3x = \int_D T^{\mu\nu}n_\mu n_\nu\sqrt{\gamma} \; d^3 x \\ \ .
\ee
By imposing the dominant energy condition on the matter energy-momentum tensor (see section
\ref{s:ADM}), the vector $-T^{\mu\nu}n_\nu$ is  future directed and non-space-like. Its 
 Lorentzian norm is therefore non-positive and an associated 
matter mass density $m$ can be given as $m^2:=-P^\mu P_\mu=
-(-T^{\mu\rho}n_\rho)(-T^{\nu\sigma}n_\sigma)g_{\mu\nu}=E^2-p^i p_i \geq 0$.
The corresponding mass $\mathrm{M}[D]$ in the extended region $D$ would be
\be
\label{e:mattermass}
\mathrm{M}[D]:=\int_D \sqrt{E^2-p^i p_i} \; \sqrt{\gamma}\;d^3x \ \ .
\ee
Note the difference between the construction of $\mathrm{M}[D]$ and that of $m[D]$
in the Minkowskian case: for the latter one first integrates to obtain the charges and then 
calculates a Minkowskian norm, whereas for constructing
$\mathrm{M}[D]$ that order is reversed; in addition, different metrics are 
employed in each case (cf. section 2.2. in \cite{Szaba04}).

\subsection{Problems when defining a gravitational energy-momentum}
\label{s:problems}
In the characterisation of the physical properties of the gravitational
field, in particular its energy-momentum and angular
momentum, we could try to follow a similar strategy to that
employed for the matter fields. This would amount to 
identify appropriate
local densities that would then be  
integrated over finite spacetime regions.
However such an approach
rapidly meets important conceptual difficulties.

A local (point-like)
density of energy associated with the gravitational field
cannot be defined in General Relativity. Reasons for this can be tracked to the Equivalence
Principle. Illustrated in a heuristic manner, this principle can
be used to get rid of the gravitational field
on a given point of spacetime. Namely, a  free falling 
point-like particle does not {\em feel} any
gravitational field so that, in particular, no gravitational
energy density can be identified at spacetime points.

In a Lagrangian setting, these basic conceptual difficulties are reflected in the
attempts to construct a gravitational energy-momentum 
tensor, when mimicking the methodological steps
followed in the matter field case.
We can write generically the gravitational-matter action as
\be
\label{e:action}
S = S_\mathrm{EH} + S_\mathrm{m} =
\frac{1}{16\pi}\int_{\cal M} {}^4\!R \sqrt{-g}\;d^4 x 
+ \int_{\cal M} L_\mathrm{m}(g_{\mu\nu}, \Phi_i, \nabla_\mu \Phi_i,...)
\sqrt{-g}\;d^4 x  \ \ .
\ee
where $ S_\mathrm{EH}$ denotes the Einstein-Hilbert action
and $\Phi_i$ in the matter Lagrangian $ L_\mathrm{m}$ account for the
matter fields.
The symmetric energy-momentum for matter is obtained 
from the variation of the matter action $S_\mathrm{m}$ with respect to the metric
\be
\label{e:symmetrucEM}
T_{\mu\nu} := \frac{-2}{\sqrt{-g}} 
\frac{\delta S_\mathrm{m}}{\delta g^{\mu\nu}} \ \ ,
\ee
whereas the field equations for the matter fields follow from the variation
with respect to the matter fields $\Phi_i$. On the contrary,
the gravitational action $S_\mathrm{EH}$ only depends on the gravitational field,
since any
further background structure would be precluded by diffeomorphism
invariance (a feature closely tied to the physical Equivalence Principle). 
Einstein equation for the gravitational field follows from the 
variation of the total action with respect to
the metric field $g_{\mu\nu}$, with no gravitational analogue
of the symmetric matter  energy-momentum tensor
$T_{\mu\nu}$. Attempts to construct a symmetric energy-momentum
tensor for the gravitational field either recover the Einstein tensor 
$G_{\mu\nu}$ or can only be related to  higher-order gravitational
energy-momentum objects, such as the Bel-Robinson tensor (see e.g. 
\cite{Senov00}). Again, the absence of a tensorial (i.e. point-like geometric) quantity
representing energy-momentum for the gravitational field is consistent with,
and actually a consequence of, the Equivalence Principle.

The natural interpretation of the symmetric matter energy-momentum tensor $T_{\mu\nu}$
as introduced in Eq. (\ref{e:symmetrucEM}) is that of the
{\em current-source} for the gravitational field, obtained as a conserved current 
associated with spacetime translations. Alternative, in terms
of the Noether theorem \cite{Noeth18} it is natural to introduce a 
(non-symmetric) {\em canonical} energy-momentum tensor for matter
from which a symmetric one can be constructed through the 
Belinfante-Rosenfeld procedure \cite{Belif39,Belif40,GreRei96}. The application 
of this construction to the gravitational field naturally
leads to the discussion of gravitational energy-momentum {\em pseudo-tensors}
\cite{Szaba04}. The underlying idea consists in decomposing the
Einstein tensor $G_{\mu\nu}$ into a part that can be identified with the
energy-momentum and a second piece that can be expressed in terms of a pseudo-potential.
That is \cite{ChaNesChe99}
\be
\label{e:pseudo-tensor-decomp}
{G_\mu}^\nu := -8\pi \; {t_\mu}^\nu
+\frac{1}{2\sqrt{-g}}\partial_\lambda({H_\mu}^{\nu\lambda}) \ \ ,
\ee
where $t_{\mu\nu}$ is the gravitational energy-momentum pseudo-tensor and
${H_{\mu\nu}}^\lambda$ is the superpotential.
Einstein equation is then written as
\be
\partial_\lambda({H_\mu}^{\nu\lambda})=
16\pi \sqrt{-g}\left({t_\mu}^ \nu+{T_\mu}^\nu  \right)=:16\pi {{\cal T}_\mu}^\nu \ \  .
\ee
Objects ${t_\mu}^\nu$ and ${H_\mu}^ {\nu\lambda}$ are not tensorial
quantities. This means that their value
at a given spacetime point is not a well-defined notion.
Moreover, their very definition needs the introduction of some additional
background structure and some 
choice of preferred coordinates is naturally involved.
Different pseudo-tensors exist in the literature, e.g. 
those introduced by Einstein, Papapetrou, Bergmann, Landau and Lifshitz,
Moller or Weinberg (e.g. see references in \cite{ChaNesChe99}).

As an alternative to the pseudo-tensor approach, there also exist attempts in the
literature aiming at constructing 
truly tensorial energy-momentum quantities. However they also involve the introduction
of some additional structure, either in the form
of a background object or by fixing a gauge in some
given formulation of General Relativity (cf. comments on the 
tetrad formalism approach in \cite{Szaba04}).

\paragraph{Non-local character of gravitational energy}
As illustrated above, crucial conceptual and practical caveats
are involved in the association of energy and angular momentum
with the gravitational field. For these reasons, one might legitimately 
consider gravitational energy and angular momentum in General Relativity
as intrinsically meaningless notions in {\em generic} situations, in such
a way that the effort to derive explicit general local expressions actually represents
an ill-defined problem (cf. remarks in \cite{MisThoWhe73} referring to the quest
for a local expression of energy in General Relativity). Having said this and 
after accepting the 
non-existence of  a {\em local} (point-like) notion of energy density
for the gravitational field,
one may also consider gravitational energy-momentum
and angular momentum as notions intrinsically associated with 
extended domains of the spacetime and then look for restricted settings
or appropriate limits where
they can be properly defined.

In fact, making a sense of the energy and angular 
momentum for the gravitational field in given regions of spacetime is extremely important 
in different contexts of gravitational physics, as it can be   
illustrated with examples coming from mathematical relativity, black hole physics,
lines of research in Quantum Gravity, or relativistic astrophysics.
From a structural point of view,
having a well-defined {\em mass positivity} result is 
crucial for  the internal consistency of the
theory, as well as for the discussion of the solutions stability. Moreover, the 
possibility of introducing appropriate positive-definite
({\em energy}) quantities is often a key step in different developments
in mathematical relativity, in particular when using variational 
principles.
In the study of the physical picture of black holes, 
appropriate notions of mass and angular momentum are employed. 
In particular, they play a key role in the formulation of black hole 
thermodynamics (e.g. \cite{Wald01}), 
a cornerstone in different approaches to Quantum Gravity.
In the context of relativistic astrophysics and numerical
relativity, the study of relativistic binary
mergers, gravitational collapse and the associated generation/propagation 
of gravitational radiation
also requires appropriate notions of energy and angular momentum 
(see e.g. \cite{JarValGou08} for a further discussion on the 
intersection between numerical and mathematical relativity).

Once the non-local nature of the gravitational energy-momentum and
angular momentum is realised, the conceptual challenge is translated into
the manner of determining the appropriate 
 physical parameters associated with the gravitational field in an extended
region of spacetime. An unambiguous answer
has been given in the case of the total mass
of an isolated system. However, the situation is much less clear
in the case of extended but finite spacetime domains.
In a broad sense, existing attempts  either
enforce some additional structure that restricts the study 
to an appropriate subset of the solution space
of General Relativity, or alternatively they  look for a 
genuinely geometric characterisation aiming at
fulfilling some expected {\em physical} requirements.
In this article we present an overview of some of the relevant existing attempts
and illustrate the kind of additional structures they involve.

\subsection{Notation}
\label{s:notation}
Before proceeding further, we set the notation, some of whose elements
have already been anticipated above. 
The signature of spacetime $({\cal M}, g_{\mu\nu})$ is chosen to be
$\mathrm{diag[}-1,1,1,1 \mathrm{]}$ and Greek letters are used for
spacetime indices in $\{0,1,2,3\}$. We denote the  Levi-Civita connection
by $\nabla_\mu$ and the volume element by
${}^4\!\epsilon=\sqrt{-g}\;dx^0\wedge dx^1\wedge dx^2\wedge dx^3$. We make 
$G=c=1$ throughout.

\subsubsection{3+1 decompositions}
In our presentation of the subject,
3+1 foliations  of spacetime  $({\cal M}, g_{\mu\nu})$
by space-like 3-slices $\{\Sigma_t\}$  will play an important role. 
Given a height-function $t$,
the time-like unit normal to $\Sigma_t$ will be denoted by $n^\mu$
and the 3+1 decomposition of the evolution vector field 
by $t^\mu=N n^\mu+\beta^\mu$, where $N$ is the lapse function and
$\beta^\mu$ is the shift vector.
The induced metric on the space-like 3-slice $\Sigma_t$ is expressed
as $\gamma_{\mu\nu}=g_{\mu\nu}+n_\mu n_\nu$, with  $D_\mu$ the
associated Levi-Civita connection and volume element
${}^3\!\epsilon=\sqrt{\gamma}dx^1\wedge dx^2\wedge dx^3$, so that
${}^3\!\epsilon_{\mu\nu\rho}= n^\sigma {}^4\!\epsilon_{\sigma\mu\nu\rho}$. 
The extrinsic curvature
of $(\Sigma_t, \gamma_{\mu\nu})$ in $({\cal M}, g_{\mu\nu})$
is defined as $K_{\mu\nu}:=-\frac{1}{2}{\cal L}_n \gamma_{\mu\nu}=
-{\gamma_\mu}^\rho \nabla_{\rho}n_\nu$. The 3+1 decomposition
of the (matter) stress-energy tensor, in terms of an Eulerian 
observer $n^\mu$ in rest with respect to the foliation  $\{\Sigma_t\}$, is
\be
\label{e3+1Tmunu}
T_{\mu\nu}=E \; n_\mu n_\nu + p_{(\mu}n_{\nu)} + S_{\mu\nu} \ \ ,
\ee
where the matter energy and momentum densities are given by 
$E:= T_{\mu\nu} n^\mu n^\nu$ and $p_\mu:= - T_{\nu\rho}n^\nu {\gamma^\rho}_\mu$,
respectively, whereas the matter stress tensor is 
$S_{\mu\nu}:= T_{\rho\sigma} {\gamma^\rho}_\mu{\gamma^\sigma}_\nu$.
Latin indices running in $\{1,2,3\}$ will be employed in expressions only 
involving objects intrinsic to space-like $\Sigma_t$ slices.

\subsubsection{Closed 2-surfaces}
Closed 2-surfaces ${\cal S}$, namely topological spheres in our discussion,
will also be relevant in the following. The normal bundle $T^\perp \!{\cal S}$
can be spanned by a time-like unit vector field $n^\mu$ and a space-like unit 
vector field $s^\mu$, that we choose to satisfy the orthogonality condition
 $n^\mu s_\mu=0$. When considering ${\cal S}$ as embedded
in a space-like 3-surface $\Sigma$, $n^\mu$ can be identified with the time-like normal
to $\Sigma$ and $s^\mu$ with the normal to ${\cal S}$ tangent to $\Sigma$.
In the generic case, $n^\mu$ and $s^\mu$ can be defined up to a boost 
transformation: $n'^\mu = \mathrm{cosh}(\eta) n^\mu + \mathrm{sinh}(\eta) s^\mu$
and   $s'^\mu = \mathrm{sinh}(\eta) n^\mu + \mathrm{cosh}(\eta) s^\mu$, with $\eta$
a real parameter. Alternatively, one can span $T_p^\perp \!{\cal S}$ at $p \in {\cal S}$
in terms of the null normals defined by the intersection
between the normal plane to ${\cal S}$ and the light-cone at the spacetime point $p$. The directions
defined by the {\em outgoing} $\ell^\mu$ and the {\em ingoing} $k^\mu$  null normals
(satisfying $k^\mu \ell_\mu = -1$) 
are uniquely determined, though it remains a boost-normalization freedom:
$\ell'^\mu= f\cdot \ell^\mu$,   $k'^\mu=\frac{1}{f}\cdot k^\mu$. 
The induced metric on ${\cal S}$ is given by: $q_{\mu\nu}=g_{\mu\nu}+k_\mu \ell_\nu
+ \ell_\mu k_\nu = g_{\mu\nu} +n_\mu n_\nu - s_\mu s_\nu = \gamma_{\mu\nu} - s_\mu s_\nu$,
the latter expression applying when ${\cal S}$ is embedded in $(\Sigma, \gamma_{\mu\nu})$.
The Levi-Civita connection associated with $q_{\mu\nu}$ will be denoted by 
${}^2\!D_\mu$ and the volume element by ${}^2\!\epsilon = \sqrt{q} dx^1\wedge dx^2$, i.e.
${}^2\!\epsilon_{\mu\nu}= n^\rho s^\sigma {}^4\!\epsilon_{\rho\sigma\mu\nu}$.
When integrating tensors on ${\cal S}$ with components normal to the sphere,
it is convenient to express the volume element as 
$dS_{\mu\nu}=(s_\mu n_\nu - n_\mu s_\nu) \sqrt{q} d^2x$ (this is just a convenient manner of 
re-expressing ${}^4\!\epsilon_{\mu\nu\rho\sigma}$ for integrating over ${\cal S}$
after a contraction with the appropriate tensor; cf. for example Eq.(\ref{e:Komar_quantity})). 

The {\em second fundamental tensor} of $({\cal S}, q_{\mu\nu})$ in $({\cal M}, g_{\mu\nu})$
is defined as ${\cal K}^\alpha_{\mu\nu}:= {q^\rho}_\mu {q^\sigma}_\nu 
\nabla_\rho {q^\alpha}_\sigma $, that can be expressed as
 ${\cal K}^\alpha_{\mu\nu}= n^\alpha \Theta^{(n)}_{\mu\nu} + 
s^\alpha \Theta^{(s)}_{\mu\nu} =  k^\alpha \Theta^{(\ell)}_{\mu\nu}+
 \ell^\alpha \Theta^{(k)}_{\mu\nu}$, where the {\em deformation tensor}
$\Theta^{(v)}_{\mu\nu}$ associated with a vector $v^\mu$ normal to ${\cal S}$ is 
defined as $\Theta^{(v)}_{\mu\nu}=  {q^\rho}_\mu {q^\sigma}_\nu \nabla_\rho v_\sigma$.
We set a specific notation for the cases corresponding to $s^\mu$ and $n^\mu$, namely 
$H_{\mu\nu}:= \Theta^{(s)}_{\mu\nu}$, the
extrinsic curvature of $({\cal S}, q_{\mu\nu})$ inside a 3-slice $(\Sigma, \gamma_{\mu\nu})$,
and $L_{\mu\nu}:= -\Theta^{(n)}_{\mu\nu}$.

Information about the extrinsic curvature of $({\cal S}, q_{\mu\nu})$ 
in $({\cal M}, g_{\mu\nu})$ is completed by the {\em normal fundamental forms}
associated with normal vectors $v^\mu$. In particular,
we define the 1-form $\Omega^{(\ell)}_\mu := k^\rho {q^\sigma}_\mu \nabla_\sigma \ell_\rho$.
This form is not invariant under a boost transformation, and transforms as 
$\Omega^{(\ell')}_\mu = \Omega^{(\ell)}_\mu + {}^2\!D_\mu \mathrm{ln}f$ in the notation above. 
Other normal fundamental forms can be defined in terms of normals
$k^\mu, n^\mu$ and $s^\mu$, but they are all related up to total derivatives.

\section{Spacetimes with Killing vectors: Komar quantities}
\label{s:Komar}
As commented above, some additional structure is needed to introduce meaningful notions of
gravitational energy and angular momentum.
Let us first consider spacetimes admitting isometries. This
represents the most straightforward generalization of the 
definition of physical parameters as conserved quantities under existing symmetries.
Requiring the presence of Killing vectors
represents our first example of the enforcement of an additional structure 
on the considered spacetime.

Given a Killing vector field $k^\mu$ in the spacetime $({\mathcal M}, g_{\mu\nu})$
and ${\mathcal S}$ a space-like closed 2-surface, let us define the Komar quantity
\cite{Komar59}
$k_{\rm K}$ as
\be
\label{e:Komar_quantity} 
k_{\rm K} := - \frac{1}{8\pi} \oint_{\Sp} \nabla^\mu k^\nu  \, 
dS_{\mu\nu} \ \  ,
\ee
(see previous section for the notation $dS_{\mu\nu}$  for the volume element on ${\mathcal S}$).
Let us consider ${\mathcal S}$ as embedded in a space-like 3-slice
$\Sigma$ and let us take a second closed 2-surface ${\mathcal S}'$ such
that either  ${\mathcal S}'$ is completely contained in ${\mathcal S}$ or vice-versa,
and let us denote by $V$ the region in $\Sigma$ contained between
${\cal S}$ and ${\cal S}'$. 
The previously defined Komar quantity $k_{\rm K}$
is then {\em conserved} in the sense that its value does not depend 
on the chosen 2-surface as long as no matter is present in the intermediate region $V$
\be
\label{Komar_invariance}
 	k_{\rm K}^{\cal S} = 2 \int_{V} 
 		\left( T_{\mu\nu} - \frac{1}{2} T g_{\mu\nu} \right)
 			n^\mu k^\nu \sqrt{\gm}\, d^3 x
 		 	+ k_{\rm K}^{{\cal S}'} \ \ ,
\ee
where $T=T_{\mu\nu} g^{\mu\nu}$.
\begin{remark}
Two important points must be stressed: a) the definition of $k_{\rm K}$ is geometric 
and therefore coordinate
independent, and b)  $k_{\rm K}$ is associated with a closed 2-surface
with no need to refer to any particular embedding in a 3-slice $\Sigma$ (in the
discussion above the latter has
been only introduced for pedagogical reasons).
\end{remark}

\subsection{Komar mass}
\label{s:Komar_mass}
Stationary spacetimes admit a time-like Killing vector field 
$k^\mu$. The associated conserved Komar quantity
is known as the Komar mass
\be 
\label{e:Komar_mass}
	M_{\rm K} := - \frac{1}{8\pi} \oint_{\Sp} \nabla^\mu k^\nu \, dS_{\mu\nu} \ \ .
\ee
This represents our first notion of mass in General Relativity. 
It is instructive
to write the Komar mass in terms of 3+1 quantities. Given a 
3-slicing $\{\Sigma_t\}$ and choosing the evolution vector 
$t^\mu = N n^\mu + \beta^\mu$ to coincide with the time-like Killing symmetry, we find
\be 
\label{e:3+1Komar_mass}
	M_{\rm K} =  \frac{1}{4\pi} \oint_{\Sp} 
	\left( s^i D_i N - K_{ij} s^i \beta^j \right)  \sqrt{q}
		\, d^2 x \ \ .
\ee

\subsection{Komar angular momentum}
\label{s:Komar_ang_mom}
Let us consider now an axisymmetric spacetime, where the axial
Killing vector is denoted by $\phi^\mu$. That is,    
$\phi^\mu$ is a space-like Killing vector whose action on ${\mathcal M}$ has compact orbits, 
two stationary points (the poles), and is normalized so that its natural affine parameter 
takes values in $[0, 2\pi)$.
The Komar angular momentum is defined as
\be  
\label{e:Komar_ang_mom}
J_{\rm K} :=  \frac{1}{16\pi} \oint_{\Sp_t} \nabla^\mu \phi^\nu \, dS_{\mu\nu} \ \ .
\ee
Note (apart from the sign choice) the factor $1/2$ with respect to the Komar 
quantity $\phi_{\rm K}$, known as
the {\em Komar anomalous factor} (it can be explained in the context
of a bimetric formalism by writing the conserved quantities in terms of an 
{\em Einstein energy-momentum flux} density that can be expressed as the sum of half the Komar 
contribution plus a second term: in the angular momentum case this second piece vanishes, whereas
for the mass case it equals half the Komar term; cf. \cite{Katz85}).
Adopting a 3-slicing adapted to axisymmetry, i.e. $n^\mu \phi_\mu =0$, we have:     
\be 
\label{e:3+1Komar_ang_mom}
	J_{\rm K} =  \frac{1}{8\pi} \oint_{\Sp} 
		K_{ij} s^i \phi^j \sqrt{q} \, d^2 x   =  
 \frac{1}{8\pi} \oint_{\Sp} \Omega_\mu^{(\ell)} \phi^\mu  \sqrt{q} \, d^2 x \ \ .
\ee

\section{Total mass of Isolated Systems in General Relativity}

\subsection{Asymptotic Flatness characterisation of Isolated Systems}
The characterisation of an isolated system in 
General Relativity aims at capturing the idea that spacetime becomes flat
when we move {\em sufficiently far} from the system, so that spacetime
approaches that of Minkowski.
However, the very notion of {\em far away} becomes problematic due
to the absence of an a priori background spacetime. In addition, 
we must consider different {\em kinds of infinities}, since we can
move away from the system in space-like and also in null directions.
Different strategies exist in the literature for the formalization 
of this asymptotic flatness idea, and not all of them are mathematically
equivalent. Traditional approaches attempt to specify the adequate
fall-off conditions
of the curvature in appropriate coordinate systems at {\em infinity}.
These approaches have the advantage of embodying the weakest
versions of asymptotic flatness. We will illustrate their use
in the discussion of spatial infinity in  section \ref{s:AsympEucl}.
However, the use of coordinate expressions in this strategy
also introduces the need of verifying the intrinsic nature of the obtained 
results, something that it is not always straightforward.
For this reason, a geometric manner of describing asymptotic flatness 
is also desirable, without relying on specific coordinates.
This has led to the conformal compactification picture, where infinity
is {\em brought} to a {\em finite distance} by an appropriate spacetime conformal 
transformation. More concretely, one
works with an unphysical spacetime $(\tilde{\cal M}, \tilde{g}_{\mu\nu})$
with boundary, such that the 
physical spacetime $({\cal M}, g_{\mu\nu})$ is conformally equivalent
to the interior of $(\tilde{\cal M}, \tilde{g}_{\mu\nu})$, i.e. 
$\tilde{g}_{\mu\nu}=\Omega^2 g_{\mu\nu}$. {\em Infinity} is captured by
the boundary  $\partial\tilde{\cal M}$ and is
characterised by the vanishing of the conformal factor, $\Omega=0$. The whole picture  
is inspired 
in the structure of the conformal
compactification of Minkowski spacetime.
The  conformal boundary is the union of different pieces, 
which are classified according
to the metric-type of the geodesics reaching their points. This
defines (past and future) null infinity $\mathscr{I}^\pm$, 
spatial infinity $i^0$ and (past and future) time-like infinity $i^\pm$,
i.e. $\partial\tilde{\cal M}=\mathscr{I}^\pm \cup i^0 \cup i^\pm $.
The conformal spacetime is represented in the so-called
Carter-Penrose diagram.
Fall-off conditions for the characterisation of asymptotic flatness are
substituted by differentiability
conditions on the fields at null and spatial infinity (isolated systems do not 
require flatness conditions on time-like
infinity).
Null infinity was introduced in the conformal picture by 
Penrose \cite{Pen73,Pen65a},  the discussion of asymptotic flatness
at spatial infinity was developed by Geroch \cite{Geroc72} and a 
unified treatment was presented in 
\cite{AshHan78,Ashte80} (see also \cite{AshteM79,Haywa03}).
We will briefly illustrate the different approaches to asymptotic flatness in the following
sections, but we refer
the reader to the existing bibliography (e.g. \cite{Fra04,Wald84}) for further details.

\subsection{Asymptotic Euclidean slices}
\label{s:AsympEucl}
The following two sections are devoted to the discussion
of conserved quantities at spatial infinity, but they also illustrate the
coordinate-based approach to asymptotic flatness.
A slice $\Sigma$ endowed with a space-like 3-metric $\gamma_{ij}$
is {\em asymptotically Euclidean} (flat), if there 
exists a Riemannian background metric $f_{ij}$ such that:
\begin{itemize}
\item[i)] $f_{ij}$ is flat, except possibly on a compact 
domain $D$ of $\Sigma$.
\item[ii)] There exists a coordinate system $(x^i)=(x,y,z)$ such that
outside $D$, $f_{ij} = \mathrm{diag}(1,1,1)$
({\em Cartesian-type coordinates}) and the variable $r:=\sqrt{x^2+y^2+z^2}$ can
take arbitrarily large values on $\Sigma$.
\item[iii)] When $r\rightarrow +\infty$
\bea
\label{e:asymp_Euclidean}
	\gm_{ij} = f_{ij} + O(r^{-1}) \ \ &,& \ \ 
	\der{\gm_{ij}}{x^k} = O(r^{-2}) \ \ ; \nonumber \\
	K_{ij} = O(r^{-2}) \ \ &,& \ \ 
	\der{K_{ij}}{x^k} = O(r^{-3}) \ \ .
\eea
\end{itemize}
Given an asymptotically flat spacetime foliated by asymptotically Euclidean slices
$\{\Sigma_t\}$, 
{\em spatial infinity} is defined by $r\rightarrow+\infty$ 
and denoted as $i^0$.

\subsubsection{Asymptotic symmetries at spatial infinity} 
As commented in the discussion of the Komar quantities, the existence of symmetries provides
a natural manner of defining physical parameters as conserved quantities. In the context 
of spatial infinity, the spacetime diffeomorphisms preserving the
asymptotic Euclidean structure (\ref{e:asymp_Euclidean}) are referred to as asymptotic symmetries. 
Asymptotic symmetries close a Lie group.
Since the spacetime is asymptotically flat, one would expect
this group to be isomorphic to the Poincar\'e group. However,
the set of  diffeomorphisms 
$(x^\mu)=(t,x^i)\rightarrow({x'}^\mu)=(t',{x'}^i)$
preserving conditions (\ref{e:asymp_Euclidean}) is given by
\be 
\label{e:asymp_sym_repres}
	{x'}^\mu = \Lambda^\mu_{\ \, \nu} x^\nu 
	+ c^\mu(\theta,\varphi) + O(r^{-1}) \ \ ,
\ee
where  $\Lambda^\mu_{\ \, \nu}$ is a Lorentz matrix and the $c^\mu$'s
are four functions of the angles $(\theta,\varphi)$ related to  coordinates
$(x^i)=(x,y,z)$ by the standard spherical formul\ae: $x = r \sin\theta\cos\varphi, y 
= r \sin\theta\sin\varphi, z = r \cos\theta$.
This group indeed contains the Poincar\'e symmetry, but
it is actually much larger due to the presence of {\em angle-dependent} translations.
The latter are known as {\em supertranslations} and are defined by 
$c^\mu(\theta,\varphi)\not={\rm const}$
and $\Lambda^\mu_{\ \, \nu} = \delta^\mu_{\ \, \nu}$
in the group representation (\ref{e:asymp_sym_repres}).
The corresponding abstract infinite-dimensional symmetry preserving the structure
of spatial infinity ({\em Spi}) is referred to as the Spi group
\cite{AshHan78,Ashte80}. 
The existence of this (infinite-dimensional)
Lie structure of asymptotic symmetries has implications
in the definition of a global physical mass, linear and angular momentum at spatial infinity
(see below).

\subsection{ADM quantities}
\label{s:ADM} 

Hamiltonian techniques are particularly powerful for the systematic
study of physical parameters, considered as conserved quantities under symmetries
acting as canonical transformations in the solution (phase)
space of a theory. In this sense, the Hamiltonian formulation of 
General Relativity
provides a natural framework for the discussion of global quantities at spatial infinity.
This was the original approach adopted by Arnowitt, Deser and Misner 
in \cite{ArnDesMis62} and we outline here the basic steps.

First, a variational problem for the class of spacetimes
we are considering must be set. For a correct formulation we need to specify: a) the dynamical
fields we are varying, b) the domain  ${\cal V}$ over which these fields are varied
together with the  prescribed value of their variations 
at the boundary $\partial {\cal V}$, and c) the action 
functional $S$ compatible with the field equations.
As integration domain ${\cal V}$ we consider the region bounded by 
two space-like 3-slices $\Sigma_{t_1}$ and  $\Sigma_{t_2}$ and 
an outer time-like tube ${\mathcal B}$.
$\Sigma_{t_1}$ and  $\Sigma_{t_2}$ can be seen as part of 
a 3-slicing $\{\Sigma_t\}$  with metric and extrinsic curvature
given by ($\gamma_{ij}, K^{ij}$), whereas  ${\mathcal B}$ has ($\chi_{\mu\nu}, P^{\mu\nu}$) as
induced metric and extrinsic curvature. 
That is, $\chi_{\mu\nu}=g_{\mu\nu} - u_\mu u_\nu$ and 
$P_{\mu\nu} = -{\gamma_\mu}^\rho \nabla_{\rho}u_\nu$, where $u^\mu$ is the unit 
space-like normal to ${\mathcal B}$.
The dynamical field whose variation we consider is the
spacetime metric $g_{\mu\nu}$, under boundary conditions 
$\delta  g_{\mu\nu}|_{\partial {\cal V}}=0$ (note that we impose nothing
on variations of the derivatives of $g_{\mu\nu}$).
The appropriate gravitational Einstein-Hilbert action then reads (cf. for example 
\cite{Poiss04}; the
discussion has a straightforward extension to incorporate matter)
\bea
\label{e:EH_action}
	S &=& \frac{1}{16\pi} \int_{\cal V} {}^4\!R \sqrt{-g} \, d^4 x  
	 + \frac{1}{8\pi}\left\{
           - \int_{\Sigma_{t_2}} (K- K_0) \sqrt{\gamma} \, d^3 x \right. \\
        &&\left. + \int_{\Sigma_{t_1}} (K- K_0) \sqrt{\gamma} \, d^3 x 
        + \int_{\mathcal B} (P- P_0) \sqrt{-\chi} \, d^3 x \right\} \nn \ \ ,
\eea 
where $K$ and $P$ are the traces of the extrinsic curvatures of 
the hypersurfaces  $\Sigma_{t_i}$ and ${\cal B}$, respectively, as 
embedded  in $({\cal M},g_{\mu\nu})$.
The subindex $0$ corresponds to their  extrinsic curvatures 
as embedded in $({\cal M},\eta_{\mu\nu})$.
The boundary term guarantees the well-posedness of the variational principle, i.e. 
the functional differentiability
of the action and the recovery of the correct Einstein field equation, under the assumed
boundary conditions for the dynamical fields.

Making use of the 3+1 fields decompositions, and considering the 
intersections 	$\Sp_t := {\cal B} \cap \Sigma_t$ between space-like 
3-slices $\Sigma_t$ and the time-like hypersurface ${\mathcal B}$, we 
can express the action (\ref{e:EH_action}) as
\be 
\label{e:S_Hilbert2}
   S =  \frac{1}{16 \pi}\int_{t_1}^{t_2} \left\{ \int_{\Sigma_t} N 
	\left({}^3\!R+K_{ij}K^{ij} -K^2\right) \sqrt{\gam} \, d^3 x 
+ 2 \oint_{\Sp_t} \left (H-H_0\right)\sqrt{q}\;d^2 x \right\}dt 
\ee
where $H$ and $H_0$ denote the trace of the extrinsic curvature 
of the 2-surface $\Sp_t$ as embedded in $(\Sigma_t, \gamma_{ij})$ and  
$(\Sigma_t, f_{ij})$, respectively. The Lagrangian density $L$ can be read 
from the form of the action (\ref{e:S_Hilbert2}).
The 3-metric $\gamma_{ij}$ plays the role of the dynamical variable 
and the dependence of $L$ on $\dot{\gamma}_{ij}$
follows from the explicit expression of the
extrinsic curvature $K_{ij}$ in terms of the lapse and the shift, that is
\be 
\label{e:dec:Kij_gam_dot}
	K_{ij} = \frac{1}{2N} \left( \gam_{ik} D_j \beta^k + 
	\gam_{jk} D_i \beta^k - {\dot\gam}_{ij} \right) \ \ . 
\ee
In particular no derivatives of  $N$ and 
$\beta^i$ appear in (\ref{e:S_Hilbert2}), indicating that the lapse function and 
the shift vector are not dynamical variables.
The Hamiltonian description is obtained by performing a 
Legendre transformation from variables $(\gamma_{ij}, \dot{\gamma}_{ij})$
to canonical ones $(\gamma_{ij}, \pi^{ij})$, where
\be
\label{e:canonical_momemtum}
\pi^{ij}:= \frac{\delta L}{\delta \dot{\gamma}_{ij}}=\frac{1}{16\pi} \sqrt{\gam}
\left(  K \gam^{ij} - K^{ij}  \right) .
\ee
The Hamiltonian density	$\mathcal{H}$  is then given by 
\be
	\mathcal{H} = \pi^{ij} {\dot\gam}_{ij} - L \ \ ,
\ee
and the Hamiltonian follows from an integration over a 3-slice, resulting in
(cf. \cite{Gou07a,Poiss04} for details)
\bea \label{e:glo:Ham}
H &=& \frac{1}{16\pi}\left\{ - \int_{\Sigma_t} \left( N C_0 + 2 \beta^i C_i \right) \sqrt{\gm} \; d^3 x 
\right.\\
&& 	\left. - 2 \oint_{\Sp_t} \left[ N(H-H_0) - \beta^i (K_{ij} - K \gm_{ij})
	s^j \right] \sqrt{q} \, d^2 x \right\} \ \ , \nn
\eea
where
\bea
	C_0 &:=& {}^3\!R+K^2-K_{ij}K^{ij} \ \ ,\nonumber \\
	C_i &:=& D_j K^j_{\ \, i} - D_i K \ \ .
\eea
Functionals $C_0$ and $C_i$ vanish on solutions of the Einstein equation (in vacuum).
More specifically, equations
$C_0 = 0$ and $C_i = 0$
respectively represent the Hamiltonian and momentum constraints of General Relativity,
corresponding to the contraction of the Einstein equation (\ref{e:Einstein}) with $n^\mu$.
From a geometric point of view, they are referred to as the 
Gauss-Codazzi relations and represent conditions for the
embedding of $(\Sigma_t, \gamma_{ij})$ as a submanifold of a
spacetime $({\mathcal M}, g_{\mu\nu})$ with vanishing $n^\mu G_{\mu\nu}$.
The evaluation of the gravitational Hamiltonian (\ref{e:glo:Ham}) 
on solutions to the Einstein equation yields
\be
\label{e:H_solution} 
	H_{\rm solution} = - \frac{1}{8\pi} \oint_{\Sp_t} \left[ N(H-H_0) 
	- \beta^i (K_{ij} - K \gm_{ij})
	s^j \right] \sqrt{q} \, d^2 x \ \ . 
\ee
\begin{remark}
Note that in the absence of boundaries the gravitational Hamiltonian
vanishes on physical solutions. This is a feature of diffeomorphism
invariant theories \cite{HenTei92} and reflects the fact that the Hamiltonian,
considered as the generator of a canonical transformation, does not move
points in the solution space of the theory. In other words, it is a generator
of gauge transformations, something consistent with the interpretation
of the Hamiltonian as the generator of diffeomorphisms. Note also that
the situation changes in the presence of boundaries, where diffeomorphisms
not preserving boundary conditions do not correspond to gauge transformations,
indicating the presence of residual degrees of freedom (this is 
 of relevance, for instance, in certain aspects of the quantum theory).
\end{remark}

\subsubsection{ADM energy}
We focus on solutions corresponding to isolated systems and consider
3-slices $\Sigma_t$ that are
asymptotically Euclidean in the sense of conditions (\ref{e:asymp_Euclidean})
(we refer the reader to \cite{AbbDes82} for a discussion of the total energy
in cosmological asymptotically Anti-de Sitter spacetimes and to
\cite{DesTek03} for its discussion in higher curvature gravity theories). 
We choose the lapse and the shift so that
the evolution vector $t^\mu$ is  
associated with some asymptotically inertial observer for which 
$N=1$ and $\beta^i=0$ at spatial infinity. In particular, this flow vector $t^\mu$
generates asymptotic time translations that, in this asymptotically flat 
context, constitute actual (asymptotic) symmetries.
Conserved quantities under time translations 
have the physical meaning of an energy. 
In the present case, the conserved quantity is referred to 
as the ADM energy. The latter is obtained
from expression (\ref{e:H_solution})
by making $N=1$ and $\beta^i=0$ and taking the limit to spatial 
infinity, namely $r\to \infty$ in the well-defined sense of 
section \ref{s:AsympEucl}. That is
\be 
\label{e:ADMenergy}
	E_{\rm ADM} := - \frac{1}{8\pi}\lim_{\Sp_{(t, r\rightarrow\infty})}
	\oint_{\Sp_t}(H-H_0) \sqrt{q} \, d^2 x  \ \ .
\ee
This ADM energy represents the total energy contained in the slice $\Sigma_t$.
Using the explicit expression of the extrinsic curvature in terms of
metric components, the ADM energy can be written as
\be \label{e:glo:M_ADM_cov}
	E_{\rm ADM} = \frac{1}{16\pi}
	\lim_{\Sp_{(t, r\rightarrow\infty})}
	\oint_{\Sp_t} \left[ \Df^j \gm_{ij} - \Df_i (f^{kl} \gm_{kl}) \right]
	s^i  \sqrt{q}\, d^2 x \ \ ,
\ee
where ${\mathcal D}_i$ stands for the connection associated with the metric $f_{ij}$
and, consistently with notation in section \ref{s:notation}, 
$s^i$ corresponds to the unit normal to $\Sp_t$ 
tangent to $\Sigma_t$ and oriented towards the exterior of $\Sp_t$ (note that when
$r\to \infty$ the normalization with respect to $\gamma_{ij}$ and $f_{ij}$ 
are equivalent). In particular, if we use the Cartesian-like coordinates 
employed in (\ref{e:asymp_Euclidean}) we recover the standard form
(see e.g, \cite{Wald84})
\be \label{e:glo:M_ADM_cart}
	E_{\rm ADM} = \frac{1}{16\pi}
	\lim_{\Sp_{(t, r\rightarrow\infty})}
	\oint_{\Sp_t} \left( \der{\gm_{ij}}{x^j} - \der{\gm_{jj}}{x^i} \right)
	s^i  \sqrt{q}\, d^2 x \ \ . 
\ee
\begin{remark}
We note that asymptotic flatness conditions (\ref{e:asymp_Euclidean}) guarantee
the finite value of the integral 
since the $O(r^2)$ part of the measure $\sqrt{q}\, d^2 x$ 
is compensated by the $O(r^{-2})$ parts of $\dert{\gm_{ij}}{x^j}$ and  
$\dert{\gm_{jj}}{x^i}$. It is very important to point out that finiteness of the ADM energy
relies on the subtraction of the {\em reference} value $H_0$ in Eq. (\ref{e:ADMenergy}).
\end{remark}

\paragraph{Conformal decomposition expression of the ADM energy.}
A useful expression for the ADM energy in certain formulations of the Einstein equation
is given in terms of a conformal decomposition of the 3-metric
\bea
\label{e:conformalmetric}
\gamma_{ij} = \Psi^4 \tilde{\gamma}_{ij} \ \ .
\eea
Choosing the representative  $\tilde{\gamma}_{ij}$ of the 
conformal class by the unimodular condition $\det(\tgm_{ij}) = \det(f_{ij}) = 1$,
conditions (\ref{e:asymp_Euclidean}) translate into
\bea 
\label{e:glo:Psi_tgm_asymp}
	\Psi = 1 + O(r^{-1}) \qquad &,& \qquad \der{\Psi}{x^k} = O(r^{-2}) \ \ ;\nn \\
	\tgm_{ij} = f_{ij} + O(r^{-1}) 
	\qquad &,& \qquad \der{\tgm_{ij}}{x^k} = O(r^{-2}) \ \ ,
\eea
for the conformal factor and the conformal metric. Then it follows \cite{Gou07a} 
\be 
\label{e:glo:M_ADM_Psi}
	E_{\rm ADM} = - \frac{1}{2\pi}
	\lim_{\Sp_{(t, r\rightarrow\infty})}
	\oint_{\Sp_t} s^i \left( \Df_i \Psi - \frac{1}{8} \Df^j\tgm_{ij} \right)
	 \sqrt{q}\, d^2 x \ \ .  	
\ee
Note that, whereas in the time-symmetric ($K_{\mu\nu}=0$) conformally flat case the Komar
mass is given in terms of the monopolar term in the asymptotic expansion
of the (adapted) lapse, the ADM energy is given by the monopolar term
in $\psi$ (the latter holds more generally under a vanishing {\em Dirac-like} gauge condition
on $\Df^j\tgm_{ij}$).

\begin{example}[Newtonian limit]
As an application of expression (\ref{e:glo:M_ADM_Psi}) we check that the ADM energy 
recovers the standard result in the Newtonian limit.
For this we assume that the gravitational field is weak and static.
In this setting it is always possible to find a coordinate system 
$(x^\mu)=(x^0=ct,x^i)$ such that the metric components take the form 
\be \label{e:mat:gab_weak}
   -d\tau^2 = g_{\mu\nu} dx^\mu dx^\nu =
	- \left( 1 + 2\Phi \right)  \, dt^2
	+ \left( 1 - 2\Phi \right) f_{ij} \, dx^i dx^j \ \ ,
\ee
where again $f_{ij}$  is the flat Euclidean metric in the 3-dimensional slice 
and $\Phi$ is the Newtonian gravitational potential, solution of the
Poisson equation $\Delta\Phi=4\pi \rho$ where $\rho$ is the mass density
(we recall that we use units in which the Newton's
gravitational constant $G$ and the light velocity $c$ are unity).
Then, using $\Psi =( 1 - 2\Phi)^{1/4}\approx 1 - \frac{1}{2}\Phi $,
Eq. (\ref{e:glo:M_ADM_Psi}) translates into 
\be 
	 E_{\rm ADM} =  \frac{1}{4\pi}
	\lim_{\Sp_{(t, r\rightarrow\infty})}
 	\oint_{\Sp_t} s^i \Df_i \Phi \, \sqrt{q}\, d^2 x  = 	
         \frac{1}{4\pi}  \int_{\Sigma_t}
		 \Delta \Phi \, \sqrt{f} \, d^3 x  \ \ . 
\ee
where in the second step we have assumed
that $\Sigma_t$ has the topology of $\mathbb{R}^3$
and have applied the Gauss-Ostrogradsky theorem (with $\Delta=\Df_i \Df^i$).
Using now that $\Phi$ is a solution of the Poisson equation, we can write
\be \label{e:glo:M_ADM_Newt}
	E_{\rm ADM} =  \int_{\Sigma_t}
		 \rho \, \sqrt{f} \, d^3 x  \ \ , 
\ee
and we recover the standard expression for the 
total mass of the system at the
Newtonian limit (as it will be seen in next section, in a non-boosted slice
like this, mass is directly given by the energy expression).
\end{example}

\subsubsection{ADM 4-momentum. ADM mass}
\paragraph{ADM linear momentum}
Linear momentum corresponds to the conserved quantity
associated with an invariance under spatial translations.
In the asymptotically flat case, the ADM momentum
is associated with space translations preserving 
the fall-off conditions (\ref{e:asymp_Euclidean}) expressed in terms
of the Cartesian-type coordinates $(x^i)$. 
Given one of such coordinate systems, the three vectors  
$(\partial_i)_{i\in\{1,2,3\}}$ represent asymptotic symmetries
generating asymptotic spatial translations that correspond
to a choice $N=0$ and $\beta_{(\partial_j)}^i= \delta^i_j$
in the evolution vector $t ^\mu$. Substituting these values
for the lapse and shift in the Hamiltonian expression
evaluated on solutions (\ref{e:H_solution}), we obtain the conserved quantity
under the infinitesimal translation $\partial_i$:
\be \label{e:glo:Pi_ADM_def}
	P_i := \frac{1}{8\pi} \lim_{\Sp_{(t, r\rightarrow\infty})}
	\oint_{\Sp_t} \left( K_{ik} - K \gm_{ik} \right)  \, 
	s^k \sqrt{q}\, d^2 x \ \ . 
\ee
\begin{remark}
Asymptotic fall-off conditions (\ref{e:asymp_Euclidean}) guarantee the finiteness of
expression (\ref{e:glo:Pi_ADM_def}) for $P_i$.
\end{remark}
The {\em ADM momentum} associated with the hypersurface $\Sigma_t$
is defined as the linear form $(P_i)=(P_1,P_2,P_3)$. 
Its components actually transform as those of a 
linear form under changes of Cartesian coordinates $(x^i)\rightarrow({x'}^i)$ which asymptotically
correspond to a rotation and/or a translation. 
For discussing transformations under 
the full Poincar\'e group, we must introduce the ADM 4-momentum defined
as
\be
	(P_\mu^{\rm ADM}) := (-E_{\rm ADM}, P_1, P_2, P_3) \ \ .
\ee
Under a 
coordinate change $(x^\mu)=(t,x^i)\rightarrow({x'}^\mu)=(t',{x'}^i)$ 
which preserves the asymptotic conditions 
(\ref{e:asymp_Euclidean}), i.e. any coordinate
change of the form (\ref{e:asymp_sym_repres}), components
$P_\mu^{\rm ADM}$ transform under the vector linear representation
of the Lorentz group
\be
	{P'}_\mu^{\rm ADM} = (\Lambda^{-1})^\nu_{\ \, \mu} \;  
	P_\nu^{\rm ADM} \ \ , 
\ee
as first shown by Arnowitt, Deser and Misner in \cite{ArnDesMis62}. Therefore 
$(P_\mu^{\rm ADM})$ can be seen as a linear form
acting on vectors at spatial infinity $i^0$ and is called the
{\em ADM 4-momentum}. 

\paragraph{ADM mass}
Having  introduced the ADM 4-momentum, its Minkowskian length 
provides a notion of mass.
The ADM mass is therefore defined as:
\be
\label{e:ADMmass}
M^2_{\rm ADM}:= - P_\mu^{\rm ADM} P^\mu_{\rm ADM} \ \ , \ \ 
M_{\rm ADM}=\sqrt{E^2_{\rm ADM} - P_i P^i} \ \ .
\ee	
\begin{remark}
In the literature, references are found where the term {\em ADM mass} actually
refers to this length of the ADM 4-momentum and other references where it refers to its 
time component,
that we have named here as the ADM energy. These differences somehow reflect traditional
usages in Special Relativity where the term {\em mass} is sometimes reserved to refer
to the  Poincar\'e invariant (rest-mass) quantity, and in other occasions is used to denote the
boost-dependent time component of the energy-momentum. 
\end{remark}
The ADM mass is a time independent quantity. Time evolution 
is generated by the Hamiltonian in expression (\ref{e:glo:Ham}). The time variation
of a given quantity $F$ defined on the phase space is expressed as the sum of its
Poisson bracket with the Hamiltonian (accounting for the implicit time
dependence through the time variation of the phase space variables)
and the partial derivative of $F$ with respect to time. 
Since in expression (\ref{e:glo:Ham}) there is no explicit time dependence, 
constancy of the ADM mass follows:
\be
\label{e:ADMconstant}
\frac{d}{dt} M_{\rm ADM}= 0 \ \ .
\ee
As a consequence of this, the ADM mass is a property of the whole (asymptotically
flat) spacetime.
\begin{remark}[Relation between ADM and Komar masses] Komar mass is defined
only in the presence of a time-like
Killing vector $k^\mu$ (more generally, cf.  \cite{Misne63} for an 
early critical account 
of its physical properties). However, in the asymptotically flat case we 
can discuss the relation between the ADM energy and the Komar mass associated with
an asymptotic inertial observer. Though the relation is not straightforward 
from explicit expressions (\ref{e:3+1Komar_ang_mom}) and (\ref{e:glo:M_ADM_cov}),  
it can be shown \cite{Beig78,AshteM79} that, for 
any foliation $\{\Sigma_t\}$ such that the associated unit normal $n^\mu$
coincides with the time-like Killing vector $k^\mu$ at infinity (i.e.
$N\to 1$ and $\beta^i\to 0$) we have
\be
	M_{\rm K} = M_{\rm ADM} \ \ . 
\ee
As a practical application, this relation has been used as a quasi-equilibrium condition
in the construction of initial data for compact objects
in quasi-circular orbits (e.g. \cite{GraGouBon02}).
\end{remark}

\paragraph{Positivity of the ADM mass.}
One of the most important results in General Relativity is the proof
of the positivity of the ADM mass under appropriate energy conditions
for the matter energy-momentum tensor. This is important first on conceptual
grounds, since it represents a crucial test of the internal consistency 
of the theory. A violation of this result would evidence an essential
instability of the solutions of the theory. It is also relevant on a practical
level, since this theorem (and/or related results) pervades
the everyday practice of (mathematical) relativists.

The theorem states that, under the {\em dominant energy condition}, the
ADM mass cannot be negative, i.e. $ M_{\rm ADM}\geq 0$. Moreover,
$ M_{\rm ADM}=0$ if and only if the spacetime is Minkowski.
This result was first obtained by Schoen and Yau 
\cite{SchYau79,SchYau81}  and then recovered using spinorial techniques
by Witten \cite{Wit81} (see in this sense \cite{DesTei77} for a previous 
related mass positivity result in supergravity).

The dominant energy condition essentially states that the local energy measured
by a causal observer is always positive, and that the flow of energy
associated with this observer cannot travel faster 
than light. More precisely, given a future-directed time-like vector $v^\mu$,
this conditions states that the vector $-{T^\mu}_\nu v^\nu$ is a future-oriented
causal vector. Vector $-{T^\mu}_\nu v^\nu$ represents the
 energy-momentum 4-current density as seen by the observer associated with $v^\mu$, 
in an analogous decomposition to that in (\ref{e3+1Tmunu}). From the
dominant energy condition it follows $E:= T_{\mu\nu}v^\mu v^\nu\geq 0$, i.e. the local
density cannot be negative ({\em weak energy condition}) and, more
generally, $E\geq \sqrt{P ^i P_i}$.

\subsubsection{ADM angular momentum}
Pushing forward the strategy followed for defining
the ADM mass and linear momentum, one would attempt to introduce
total angular momentum as the conserved quantity associated with 
rotations at spatial infinity. More specifically, in the Cartesian-type
coordinates used for characterising asymptotically Euclidean slices
(\ref{e:asymp_Euclidean}), infinitesimal
generators $(\phi_i)_{i\in\{1,2,3\}}$ for rotations around the three spatial 
axes are 
\bea
	\phi_x = - z \partial_y + y \partial_z  \ \ , \ \ 
	\phi_y = - x \partial_z + z \partial_x  \ \ , \ \  
	\phi_z = - y \partial_x + x \partial_y \label{e:glo:rot_flat_z} \ \ ,
\eea
which constitute Killing symmetries of the asymptotically flat metric.
When using the associated {\em lapse} functions and {\em shift} vectors
in the Hamiltonian expression (\ref{e:H_solution}), namely $N=0$
and $\beta^i_{(\phi_j)} = (\phi_j)^i$, the following three quantities result
\be 
\label{e:glo:angu_mom_def}
 J_i := \frac{1}{8\pi} \lim_{\Sp_{(t, r\rightarrow\infty})}
	\oint_{\Sp_t} \left( K_{jk} - K \gm_{jk} \right) (\phi_i)^j \,
	s^k \sqrt{q}\, d^2 x, 
\qquad i\in\{1,2,3\} \ \ .
\ee
However, the interpretation of $J_i$ as the components of an
angular momentum faces two problems:
\begin{enumerate}
\item First, asymptotic fall-off conditions (\ref{e:asymp_Euclidean}) are not sufficient
to guarantee the finiteness of expressions (\ref{e:glo:angu_mom_def}).
\item Second, in contrast with the linear momentum case, the quantity
$(J_i) = (J_1,J_2,J_3)$ does not transform appropriately under 
transformations (\ref{e:asymp_sym_repres})  
preserving (\ref{e:asymp_Euclidean}). This can be tracked to the
existence of supertranslations. 
In particular, the so-defined angular-momentum vector $(J_i)$ depends non-covariantly
on the particular coordinates we have chosen.
\end{enumerate}
For this reason, it is not appropriate to refer to an ADM angular momentum
in the same sense that we use the ADM term for mass and linear momentum quantities.
A manner of removing the above-commented ambiguities consists in identifying an
appropriate subclass of Cartesian-type coordinates where, first,
the $J_i$ components are finite and, second,
they transform as the components of a linear form.
Among the different strategies proposed in the literature, 
we comment here on the one proposed by York \cite{York79} in terms of further conditions on
the conformal metric $\tilde{\gamma}_{ij}$ introduced
in (\ref{e:conformalmetric}) and the trace of the extrinsic curvature $K$.  Namely
\bea 
	\der{\tgm_{ij}}{x^j} = O(r^{-3})  \ \ , \ \
	K = O(r^{-3}) \ \ , \label{e:glo:gauge}
\eea
representing {\em asymptotic gauge conditions}. That is, they actually impose restrictions
on the choice of coordinates but not on the geometric properties of spacetime
at spatial infinity. First condition in (\ref{e:glo:gauge}) is known as the 
{\em quasi-isotropic gauge}, 
whereas the second one is referred to as the {\em asymptotic maximal gauge}.
\begin{remark}
Note that, in contrast with the total angular momentum defined at spatial infinity, no
ambiguity shows up in the definition of the Komar angular momentum in 
Eq. (\ref{e:Komar_ang_mom}).
\end{remark}

\subsection{Bondi energy and linear momentum}
We could introduce Bondi (or Trautman-Bondi-Sachs) energy
at null infinity following the same approach 
we have employed for the ADM energy, i.e. by taking the
appropriate limit of (\ref{e:H_solution}) 
with $N=1$ and $\beta^i=0$. In the present case, instead of keeping
$t$ constant and making $r\to\infty$ as we did in (\ref{e:ADMenergy}), 
we should introduce retarded and advanced time coordinates
(respectively, $u=t-r$ and $v=t+r$) and consider the limit 
\be
\label{e_Bondienergy1}
	E_{\rm BS} := - \frac{1}{8\pi}\lim_{\Sp_{(u, v\rightarrow\infty})}
	\oint_{\Sp_u}(H-H_0) \sqrt{q} \, d^2 x \ \ .
\ee
The full discussion of this limit would require
the introduction of the appropriate fall-off conditions
for the metric components in a special class of coordinate
system adapted to null infinity (Bondi coordinates). This is in the spirit
of the original discussion on the energy flux of gravitational radiation
from an isolated system 
by Bondi, Van der Burg and Metzner \cite{BonBurMet62}, and Sachs \cite{Sac62c}.
However, aiming at providing some flavour of the geometric approach to 
asymptotic flatness, we rather outline here a discussion in the 
setting of the conformal compactification approach.

\subsubsection{Null infinity}
A smooth spacetime $({\cal M}, g)$ is {\em asymptotically simple} 
\cite{Pen63} (see e.g. also \cite{Fra04})
if  there exists another (unphysical) smooth Lorentz manifold $(\tilde{\cal M}, \tilde{g})$
such that:
\bit
\item[i)] ${\cal M}$ is an open submanifold of 
 $\tilde{\cal M}$ with (smooth) boundary $\partial\tilde{\cal M}$.
\item[ii)] There is a smooth scalar field $\Omega$ on $\tilde{\cal M}$, 
such that: $\Omega>0$, $\tilde{g}_{\mu\nu}= \Omega^2 g_{\mu\nu}$ on ${\cal M}$, and  
$\Omega=0$, $\partial_\mu\Omega\neq 0$ on $\partial\tilde{\cal M}$.
\item[iii)] Every null geodesic in  ${\cal M}$ begins and ends on 
$\partial\tilde{\cal M}$. 
\eit
An asymptotically simple spacetime is {\em asymptotically flat} 
(at null infinity) if, 
in addition,  Einstein vacuum equation is satisfied 
in a neighbourhood of $\partial\tilde{\cal M}$ (or the energy-momentum decreases
sufficiently fast in the matter case).
In this case the boundary $\partial \tilde{\cal M}$ consists,
at least, of a null hypersurface with two connected
components $\mathscr{I}= \mathscr{I}^-\cup 
\mathscr{I}^+$, each one with topology $S^2\times \mathbb{R}$
(note that in Minkowski $\partial\tilde{\cal M}$ also contains the
points $i^0, i^\pm$). Boundaries $\mathscr{I}^-$ and $\mathscr{I}^+$ 
represent past and future null infinity, respectively.

\subsubsection{Symmetries at null infinity}
In order to characterise a vector $\xi^\mu$ in ${\cal M}$
as an infinitesimal asymptotic symmetry
at (future) null infinity $\mathscr{I}^+$, we must assess the vanishing
of ${\cal L}_\xi g_{\mu\nu}$ as one gets to $\mathscr{I}^+$.
For this, we require first that $\xi^\mu$, considered
as a vector field in the unphysical spacetime (i.e. under the 
immersion of ${\cal M}$ into  $\tilde{\cal M}$), can be 
smoothly extended to  $\mathscr{I}^+$. Then $\xi^\mu$ is characterised
as an asymptotic symmetry by demanding that
$\Omega^2 {\cal L}_\xi g_{\mu\nu}$ can also
be smoothly extended to $\mathscr{I}^+$ and vanishes there, that is
\be
\label{e:conformalKilling}
\left.\left(\tilde{\nabla}_{\mu}\xi_{\nu}+
\tilde{\nabla}_{\nu}\xi_{\mu}-2\Omega^{-1}\xi^\rho \tilde{\nabla}_\rho \Omega
\; \tilde{g}_{\mu\nu}\right) \right|_{\mathscr{I}^+}=0 \ \ .
\ee
Two vector fields $\xi^\mu$ and $\xi'^\mu$ are considered to generate
the same infinitesimal asymptotic symmetry if their extensions
to $\mathscr{I}^+$ coincide. The equivalence class of such vector
fields, that we will still denote by $\xi^\mu$,  generates the 
asymptotic symmetry group at $\mathscr{I}^+$. This is known as
the Bondi-Metzner-Sachs (BMS) group and is universal in the sense
that it is same for every 
asymptotically flat spacetime.
The BMS group is infinite-dimensional, as it was the case of the Spi group at spatial infinity. 
It does not only contain the Poincar\'e group, but actually 
is a semi-direct product of the Lorentz group and the infinite-dimensional
group of {\em angle dependent} supertranslations (see details 
in e.g. \cite{Wald84}). The key point for the present discussion
is that it possesses a unique {\em canonical} set of asymptotic
4-translations
characterised as the only 4-parameter subgroup of the 
supertranslations that is a {\em normal} subgroup of the BMS group.
This leads us to the Bondi-Sachs 4-momentum.

\subsubsection{Bondi-Sachs 4-momentum}
As mentioned above, the original introduction of the Bondi energy
was based in the identification of certain
expansion coefficients in the line element of radiative spacetimes
in adapted (Bondi) coordinates \cite{BonBurMet62}. 
A Hamiltonian analysis, counterpart of the approach adopted in section \ref{s:ADM}
for introducing the ADM mass, can be found in \cite{WalZou99}. Here we 
rather follow a construction based on the Komar mass expression.
Though Eq. (\ref{e:Komar_quantity}) only defines a conserved
 quantity for a Killing vector $k^\mu$, the vector
fields $\xi_ a^\mu$ ($a\in\{0,1,2,3\}$)
corresponding to the 4-translations at 
$\mathscr{I}^+$ get closer to an infinitesimal
symmetry as one approaches $\mathscr{I}^+$. Therefore, one
can expect that a Komar-like expression makes sense for a given
cross-section $S_u$ of  $\mathscr{I}^+$. This is indeed
the case and the evaluation of the integral does not
depend on how we get to  $S_u$. However, the integral does depend
on the representative $\xi^\mu$ in the class of vectors corresponding to
the asymptotic symmetry. This is cured by imposing 
a divergence-free condition on $\xi^\mu$ \cite{GerWin81}.
Bondi-Sachs 4-momentum at $S_u\subset \mathscr{I}^+$ is then defined as
\be
\label{e:BS4mom}
P_a^{\rm BS} := - \frac{1}{8\pi}\lim_{(\Sp\to\Sp_u)}
		\oint_{\Sp} \nabla^\mu \xi_a^\nu dS_{\mu\nu} \ \ , \ \ 
\nabla_\mu  \xi_a^\mu = 0  \ \ .
\ee
Alternatively, ambiguities in the Komar integral can
be solved by dropping the condition on the divergence and
adding a term $\alpha\nabla_\mu \xi^\mu_a$ to the surface integral. 
When $\alpha=1$ the resulting integral is called the {\em linkage} \cite{WinTam65}.
The discussion of Bondi-Sachs angular momentum is more delicate.
We refer the reader to the discussion in section 3.2.4 of \cite{Szaba04}.

\paragraph{Bondi energy and positivity of gravitational radiation energy}
Bondi energy $E_{\mathrm{BS}}$ (the zero component
of the Bondi-Sachs 4-momentum) is a decreasing function
of the retarded time. More concretely, Bondi energy satisfies a {\em loss
equation} 
\be
\label{e:BSenergydecrease}
\frac{d E_{\mathrm{BS}} }{du}= - \int_{\Sp_u} F \sqrt{q}\;d^2x \ \ ,
\ee
where $F\geq 0$ can be expressed in terms of the squares of the so-called
{\em news} functions. In \cite{AshteM79} it is shown that, 
if the {\em news} tensor
satisfies the appropriate conditions, then Bondi mass coincides initially
with the ADM mass (see also \cite{Haywa03}). 
Bondi energy is interpreted as the remaining of
the ADM energy in the process of energy extraction by gravitational radiation. 
As for the ADM mass, a positivity result holds for the Bondi mass 
\cite{SchYau82,HorPer82}. These properties constitute the underlying 
conceptual/structural justification of our understanding of energy
radiation by gravitational waves: gravitational radiation carries
positive energy away from isolated radiating systems, and the total
radiated energy cannot be bigger than the original total ADM energy.

\section{Notions of mass for bounded regions: quasi-local masses}
\label{s:quasi-local}
As commented in section \ref{s:problems}, 
the convenience of associating energy-momentum with the gravitational
field in given regions of the spacetime is manifest in very different 
contexts of gravity physics. More specifically, mathematical and numerical
General Relativity or approaches to Quantum Gravity provide examples
where we need to associate such an energy-momentum with a {\em finite} region
of spacetime. This can be either motivated by the need to define appropriate
physical/astrophysical quantities, or by the convenience of finding quasi-local
quantities with certain desirable  
mathematical properties (e.g. positivity, monotonicity...) in the study
of a specific problem.

There exist many different approaches for introducing quasi-local prescriptions for the
mass and angular momentum.
Some of them can be seen as {\em quasi-localizations}
of successful notions for the physical parameters of the total system, such as the ADM
mass, whereas other attempts constitute genuine {\em ab initio}
methodological constructions, mainly based on Lagrangian
or Hamiltonian approaches. 
An important drawback of most of them in the context of the
present article is that, 
typically, they involve {\em constructions} that are 
difficult to capture in short mathematical definitions 
without losing the underlying physical/geometrical insights.
An excellent and comprehensive review is reference \cite{Szaba04} 
by Szabados.

\paragraph{Ingredients in the quasi-local constructions.}
First, the relevant bounded spacetime domain must be identified.
Typically, these are compact space-like domains $D$
with a boundary given by a closed 2-surface ${\cal S}$. 
Explicit expressions, such as relevant associated integrals, are
formulated in terms of either the (3-dimensional) domain $D$ itself or
on its boundary ${\cal S}$.
In particular, conserved-current strategies  permit to pass 
from the 3-volume integral to a conserved-charge-like 2-surface integral.
In other cases, 2-surface integrals are a consequence of the need of including
boundary terms for having a correct variational formulation
(as it was the case in the Hamiltonian formulation of section \ref{s:ADM}). 

We have already presented an example of quasi-local quantity in section
\ref{s:Komar}, namely the Komar
quantities. Since symmetries will be absent in the generic case,
an important ingredient in most quasi-local constructions is the 
prescription of some vector field that plays the role 
that infinitesimal symmetries had played in case of being present.
In connection with this, 
one usually needs to introduce some background
structure that can be interpreted as a kind of gauge choice.

Finally, different {\em plausibility} criteria for the assessment 
of the proposed quasi-local expressions (e.g. positivity, monotonicity,
recovery of known limits...) need to be considered (see \cite{Szaba04}).

\subsection{Some relevant quasi-local masses}

\subsubsection{Round spheres. Misner-Sharp energy}
In some special situations, as it is the 
case of isolated systems above and some exact solutions, there is 
agreement on the form of the gravitational field energy-momentum.
Another interesting case is that of spherically symmetric spacetimes, where
the rotation group $SO(3)$ acts transitively as an isometry. 
Orbits under this rotation group are {\em round} spheres ${\cal S}$.
Then, using the areal radius $r_A$ as a coordinate 
($4\pi r_A^2 = A$), an appropriate notion of mass/energy
was given by Misner and Sharp \cite{MisSha64}
\be
\label{e:Misner-Sharp}
E({\cal S}) :=\frac{1}{8}r_A^3 R_{\mu\nu\rho\sigma}{}^2\!\epsilon^{\mu\nu} \;
{}^2\!\epsilon^{\rho\sigma} \ \ ,
\ee
where  ${}^2\!\epsilon_{\mu\nu}= n^\rho s^\sigma {}^4\!\epsilon_{\rho\sigma\mu\nu}$  
(cf. section \ref{s:notation}) is
the volume element on ${\cal S}$.
This expression is related to the so-called Kodama vector $K^\mu$, that
can be defined in spherically symmetric spacetimes and such that
$\nabla_\mu(G^{\mu\nu}K_\nu)=0$. The current $S^\mu=G^{\mu\nu}K_\nu$
is thus conserved and, taking $D$ as a solid ball of radius $r_A$, 
the flux of $S^\mu$ through the round boundary $\partial D$ actually
equals the change in time of the mass expresion (\ref{e:Misner-Sharp}). 
Misner-Sharp proposal is considered as the {\em standard
form} of quasi-local mass for round spheres.

\subsubsection{Brown-York energy}
The rationale of the approach in Ref. \cite{BroYor93} to quasi-local energy
strongly relies on the 
well-posedness of a variational problem for the gravitational
action. The adopted variational formulation is essentially the one outlined 
in section \ref{s:ADM} (where the discussion was in fact based
in the treatment in \cite{Poiss04} adapted from \cite{BroYor93}).
However, if the main interest is placed in the expressions of quasi-local 
parameters and not in the details of the symplectic geometry 
of the system phase space, a full Hamiltonian analysis does not need to
be undertaken and one can rather follow a Hamilton-Jacobi one.
The latter starts from action (\ref{e:EH_action}) defined
on the spacetime domain ${\cal V}$. We recall that the boundary
$\partial{\cal V}$ is given by two space-like hypersurfaces $\Sigma_1$ and 
$\Sigma_2$ and a time-like tube ${\cal B}$, such that the 2-spheres
${\cal S}_i$ are the intersections between 
$\Sigma_i$ and ${\cal B}$.
The metric and extrinsic curvatures on $\Sigma_i$ are given
by $\gamma_{\mu\nu}$ and $K^{\mu\nu}$, whereas those on  ${\cal B}$
are denoted by $\chi_{\mu\nu}$ and $P^{\mu\nu}$. A Hamilton-Jacobi
principal function can then be introduced by evaluating
the action $S$ on classical trajectories. An arbitrary function $S^0$ of the
data on the boundaries can be added to $S$ [it is the responsible 
of the {\em reference} terms with subindex $0$ in expression (\ref{e:EH_action})].
The principal function is given by $S_{\mathrm{Cl}}:= \left(S - S^0\right)|_{\mathrm{Cl}}$
and Hamilton-Jacobi equations are obtained from its variation with respect to the data at the 
final slice $\Sigma_2$.
One of the Hamilton-Jacobi equations leads to the definition of 
a {\em surface stress-energy-momentum} tensor as
\be
\label{e:BYstress-energy}
\tau^{\mu\nu}:= \frac{-2}{\sqrt{-\chi}}\frac{\delta S_{\mathrm{Cl}}}{\delta \chi_{\mu\nu}}
= \frac{1}{8\pi}\left\{( P\chi^{\mu\nu} - P^{\mu\nu})  -  
(P_0\chi^{\mu\nu} -  P_0^{\mu\nu})\right\} \ \ .
\ee
This tensor satisfies a conservation-like equation
with a source given in terms of the matter energy-momentum tensor $T^{\mu\nu}$.
This motivates the definition of the {\em charge} $Q_{\cal S}(\xi^\mu)$ associated with a
vector $\xi^\mu$ as
\be
\label{e:BYcharge}
Q_{\cal S}(\xi^\mu):=\oint_{\cal S} \xi_\rho \tau^{\rho\nu} n_\nu \sqrt{q}\;d^2x \ \ ,
\ee
whose change along the tube ${\cal B}$  is given by a matter flux.
This expression is analogous to (\ref{e:conservedcharges}) in the matter case (here 
${\cal S}\subset {\cal B} $ 
and $n^\mu$ is the time-like unit normal to ${\cal S}$ and tangent to ${\cal B}$).

Using the 2+1 decomposition induced by a 3+1 
space-like slicing $\{\Sigma_t\}$, we can decompose the tensor $\tau^{\mu\nu}$
as we did for the matter energy-momentum tensor $T^{\mu\nu}$ in Eq. (\ref{e3+1Tmunu}).
Writing explicitly the time-like components, it results
\bea
\varepsilon&:=& n_\mu n_\nu \tau^{\mu\nu}=-\frac{1}{8\pi}(H-H^0) \ \ ,\nn \\
j_\mu&:=&-q_{\mu\nu}n_\rho  \tau^{\nu\rho} = 
\frac{1}{8\pi} q_{\mu \nu} s_\rho \left.(K\gamma^{\nu\rho} - K^{\nu\rho}
)\right|^\mathrm{Cl}_0 \label{e:BYepsilonj} \ \ .
\eea
Expressing the vector $\xi^\mu$ in the 3+1 decomposition $\xi^\mu = \xi n^\mu + \xi_\perp^\mu$
and considering a 2-surface ${\cal S}$ lying in a slice of $\{\Sigma_t\}$, we have
\be
\label{e:BYcharge2+1}
Q_{\cal S}(\xi^\mu)=\oint_{\cal S} \xi_\rho \tau^{\rho\nu} n_\nu \sqrt{q}\;d^2x =
\oint_{\cal S} \left( \xi \varepsilon - \xi_\perp^\rho j_\rho \right)\sqrt{q}\;d^2x \ \ .
\ee
The Brown-York energy is then [cf. with the ADM mass
expression (\ref{e:ADMmass})] 
\be
\label{e:BYmass}
E_{\mathrm{BY}}({\cal S}, n^\mu):=
Q_{\cal S}(n^\mu) = 
 -\frac{1}{8\pi}\oint_{\cal S} (H-H_0)\sqrt{q}\;d^2x \ \ .
\ee
Note that this expression explicitly depends on the manner in which 
${\cal S}$ is inserted in some space-like 3-slice. In this sense, it corresponds
to an energy (depending on a boost) rather than a mass.

\paragraph{Kijowski, Epp, Liu-Yau and Kijowski-Liu-Yau expressions}
We briefly comment on some expressions that can be related 
to the Brown-York energy. Studying more general 
boundary conditions than the ones in \cite{BroYor93},
Kijowski proposed the following quasi-local expression for the mass \cite{Kijow97}
\be
\label{e:Kijowskienergy}
E_{\mathrm{Kij}}:=\frac{1}{16\pi}\oint_{\cal S}\frac{(H_0)^2 - (H^2 -L^2)}{H_0}
\sqrt{q}\;d^2x \ \ ,
\ee
where  $H = H_{\mu\nu}q^{\mu\nu}$ and $L = L_{\mu\nu}q^{\mu\nu}$ are the traces 
of the extrinsic curvatures
of ${\cal S}$ with respect to unit orthogonal space-like $s^\mu$ and time-like
$n^\mu$ vectors, i.e. $n^\mu s_\mu =0$ (cf. notation in section \ref{s:notation}). 
Apart from the choice
of the background terms $H_0$, this expression 
only depends on ${\cal S}$, and not in the manner of embedding it 
into some space-like hypersurface. Using a different set of  boundary conditions,
another quasi-local quantity was introduced by Kijowski 
(referred to as a {\em free energy}). The 
same quantity was later derived by Liu and Yau, using a 
different approach \cite{LiuYau03}. We will refer to the resulting quasi-local energy 
as the Kijowski-Liu-Yau energy, having the form
\be
\label{e:KBLenergy}
E_{\mathrm{KLY}}:=\frac{1}{8\pi}\oint_{\cal S}\left(H^0- \sqrt{H^2 -L^2}\right) \ \ .
\ee
On the other hand, aiming at removing the dependence of Brown-York energy
on the space-like hypersurface,
Epp \cite{Epp00} proposed the following boost-invariant expression 
\be
\label{e:Eppmass}
E_{\mathrm{E}}:=\frac{1}{8\pi}\oint_{\cal S}\left(\sqrt{(H^0)^2 -(L^0)^2}
- \sqrt{H^2 -L^2}\right) \ \ .
\ee
Note that Brown-York energy can be seen as a gravitational field  version of
the quasi-local matter energy (\ref{e:matterenergy}), whereas Epp's expression
rather corresponds to the matter mass (\ref{e:mattermass}). For further recent
work along this approach to quasi-local mass, see \cite{WanYau09,MurTunXie09}.

\subsubsection{Hawking, Geroch and Hayward energies}
\paragraph{Hawking energy.}  
Given a topological sphere ${\cal S}$,
its Hawking energy is defined as \cite{Hawki68}
\be
\label{e:Hawking}
E_{\mathrm{H}}({\cal S})= \sqrt{\frac{A({\cal S})}{16\pi}}\left(1+\frac{1}{8\pi}
\oint_{\cal S} \theta_+ \theta_- \right)\sqrt{q}\;d^2 x \ \ ,
\ee
where $ \theta_+=q^{\mu\nu} \Theta_{\mu\nu}^{(\ell)}$ and  
$\theta_-=q^{\mu\nu} \Theta_{\mu\nu}^{(k)}$ are the expansions associated with
outgoing and ingoing null normals (cf. notation in section \ref{s:notation}).  
It can be motivated by understanding the mass surrounded
by the 2-sphere ${\cal S}$ as an estimate of the bending of 
ingoing at outgoing light rays from ${\cal S}$. An average, 
boost-independent measure
of this convergence-divergence behaviour of  light rays
is given by $\oint_{\cal S}  \theta_+ \theta_- {}^2\!\epsilon$. Then, from the
Ansatz $A+B \oint_{\cal S}  \theta_+ \theta_- {}^2\!\epsilon$, the constants $A$ and $B$
are fixed from round spheres in Minkowski and from the horizon sections in
Schwarzschild spacetime.

Hawking energy depends only on the surface ${\cal S}$
and not on any particular embedding of it in a 
space-like hypersurface.  
In the spherically symmetric case it recovers 
the standard Misner-Sharp energy (\ref{e:Misner-Sharp}). For apparent horizons,
or more generally for marginally trapped surfaces, it reduces
to the {\em irreducible mass} accounting for the energy
that cannot be extracted from a black hole by a Penrose process and
that is given entirely in terms of the area.
Hawking energy does not satisfy a positivity criterion, since it can
be negative even in Minkowski spacetime.
However, 
for large spheres approaching null infinity, $E_\mathrm{H}({\cal S})$ recovers 
Bondi-Sachs energy,
whereas for spheres approaching spatial infinity it tends to the ADM
energy. Though it is not monotonic in the generic case, monotonicity can be proved
for sequences of spheres obtained from appropriate geometric
flows.
This has a direct interest for the extension of Huisken \& Ilmanen
proof \cite{HuiIlm01} of the Riemaniann Penrose inequality to the general case.
 
\paragraph{Geroch energy.} 
For a surface ${\cal S}$
embedded in a space-like hypersurface $\Sigma$,
Geroch energy \cite{Geroc73} is defined as 
\be
\label{e:Geroch}
E_{\mathrm{G}}({\cal S}):= 
\frac{1}{16\pi}\sqrt{\frac{A({\cal S})}{16\pi}}
\oint_{\cal S} \left(2\; {}^2\!R - H^2 \right)\sqrt{q}\;d^2 x \ \ ,
\ee
where $H$ is again the trace of the extrinsic curvature of ${\cal S}$
inside $\Sigma$.
Geroch energy is never larger
than Hawking energy, but it can be proved that 
it also tends to the ADM mass for spheres approaching spatial infinity.

The relevance of Geroch energy lies on its key role in the
first proof of the Riemaniann Penrose, by Huisken \& Ilmanen \cite{HuiIlm01}
(see also section \ref{s:Penrose}).
In particular, use is made of the monotonicity
properties of $E_{\mathrm{G}}$ under an {\em inverse mean curvature}
flow in $\Sigma$.

\paragraph{Hayward energy.}
Some generalizations of Hawking energy exist. A vanishing expression
for flat spacetimes can be obtained by considering the modified expression
\be
\label{e:Hawkingmodif}
E'_{\mathrm{H}}({\cal S})= 
\sqrt{\frac{A({\cal S})}{16\pi}}\left(1+\frac{1}{8\pi}
\oint_{\cal S} \theta_+ \theta_- - 
\frac{1}{2} \sigma^+_{\mu\nu} \sigma_-^{\mu\nu}  \right)\sqrt{q}\;d^2 x \ \ ,
\ee
where the shears $\sigma^+_{\mu\nu}$  and $ \sigma^-_{\mu\nu}$ are the traceless parts of 
 $\Theta_{\mu\nu}^{(\ell)}$ and  $\Theta_{\mu\nu}^{(k)}$, respectively.
$E'_{\mathrm{H}}$ still asymptotes to the ADM energy at spatial infinity,
but does not recover Bondi-Sachs energy at null infinity (but rather Newman-Unti one; cf. references in
\cite{Szaba04}).
Related to this modified Hawking energy,
Hayward has proposed \cite{Haywa94b} another quasi-local energy expression
by taking into account the anholonomicity form $\Omega_\mu$,
one of the normal fundamental 1-forms introduced in section
\ref{s:notation}
\be
\label{e:Hayward}
E_{\mathrm{Hay}}({\cal S})= 
\sqrt{\frac{A({\cal S})}{16\pi}}\left(1+\frac{1}{8\pi}
\oint_{\cal S} \theta_+ \theta_- - \frac{1}{2} 
\sigma^+_{\mu\nu} \sigma_-^{\mu\nu}  - 2 \Omega_\mu\Omega^\mu  \right) 
\sqrt{q}\;d^2 x\ \ .
\ee
Though the divergence-free part of $\Omega_\mu$ can be related to angular momentum 
(see below), this 1-form 
is a gauge dependent object changing by a total differential
under a boost transformation. Therefore, some natural gauge for fixing the boost 
freedom is needed.

\subsubsection{Bartnik mass}
Bartnik quasi-local mass is an example of {\em quasi-localization}
of a global quantity, in particular the ADM mass. 
In very rough terms, the idea in Bartnik's construction 
consists in defining the mass of a compact space-like
3-domain $D$ as the ADM mass of that asymptotically Euclidean
slice $\Sigma$ that contains $D$ without any other {\em source} of energy.
The strategy to address this absence of further
energy is to consider all plausible extensions of $D$ into
Euclidean slices, calculate the ADM mass for all them, and then
consider the infimum of this set of ADM masses. In more 
precise terms, let us consider a compact, connected 3-hypersurface $D$
in spacetime, with
boundary ${\cal S}$  and induced metric $\gamma_{ij}$.
Bartnik's construction actually focuses on time-symmetric $K_{ij}=0$ domains $D$.
Let us also assume that a dominant energy condition (though the original formulation in 
\cite{Bartn89} makes use of a weak-energy-constraint condition) is satisfied.
In a time-symmetric context this amounts to the 
positivity of the Ricci scalar, ${}^3\!R\geq 0$. 
One can then define ${\cal P}(D)$ as 
the set of  Euclidean time-symmetric initial data sets 
$(\Sigma, \gamma_{ij})$ satisfying the dominant energy condition,
with a single asymptotic end, finite ADM mass  $M_{\mathrm{ADM}}(\Sigma)$,
not containing horizons (minimal surfaces in this context)  and extending
$D$ through its boundary ${\cal S}$. Then, Bartnik's mass \cite{Bartn89} is defined as
\be
\label{e:Bartnik_massD}
M_\mathrm{B}(D):= \mathrm{inf}\left\{ M_{\mathrm{ADM}}(\Sigma), \hbox{ such that } 
\Sigma \in {\cal P}(D) \right\} \ \ .
\ee
The {\em no-horizon} condition is needed to avoid extensions 
$(\Sigma, \gamma_{ij})$ with arbitrarily small ADM mass.
There is also a spacetime version of Bartnik's construction, not 
relying on an initial data set on $D$ but only on the geometry
of 2-surfaces ${\cal S}$. 
Let us define ${\cal P}({\cal S})$ as the set of globally
hyperbolic spacetimes $({\cal M}, g_{\mu\nu})$ satisfying the
dominant energy condition, admitting
an asymptotically  Euclidean Cauchy hypersurface $\Sigma$ with finite ADM mass,
not presenting an event horizon and such that 
${\cal S}$ is embedded (i.e. both its intrinsic and extrinsic geometry) 
in $({\cal M}, g_{\mu\nu})$. Then, one defines
\be
\label{e:Bartnik_massS}
M_\mathrm{B}({\cal S}):= \mathrm{inf}\left\{ M_{\mathrm{ADM}}({\cal M}), \hbox{ such that } 
{\cal M} \in {\cal P}({\cal S}) \right\} \ \ .
\ee
The comparison between $M_\mathrm{B}(D)$ and $M_\mathrm{B}(\partial D)$ is not
straightforward, due to issues regarding the horizon characterisation.
From the positivity of the ADM mass it follows the non-negativity of 
the Bartnik mass $M_\mathrm{B}(D)$. In fact, $M_\mathrm{B}(D)=0$ characterises $D$ as locally flat. 
From the definition (\ref{e:Bartnik_massD}) it also 
follows the monotonicity of $M_\mathrm{B}(D)$, i.e. if $D_1\subset D_2$ then 
$M_\mathrm{B}(D_1)\leq M_\mathrm{B}(D_2)$. Bartnik mass tends to the ADM mass, as 
domains $D$ tend to Euclidean slices (the proof makes use of the
Hawking energy introduced above). Another interesting feature,
consequence of the proof of the Riemannian Penrose conjecture \cite{HuiIlm01},
is that Bartnik mass reduces to the {\em standard form} $E({\cal S})$ in
Eq. (\ref{e:Misner-Sharp}) for round spheres.
However, the explicit calculation of the Bartnik mass is problematic.
An approach to its practical computability is provided by Bartnik's conjecture
stating that the infimum in (\ref{e:Bartnik_massD}) is actually a minimum
realised by an element in ${\cal P}(D)$ characterised by its 
stationarity outside $D$. 
Further developments 
of these ideas have been proposed by Bray (cf. \cite{BraChr04}).

Before concluding this subsection, we mention the explicit construction 
in \cite{YanYon09} of quantum analogs  
for some of the previous quasi-local gravitational energies (specifically
for Brown-York, Liu-Yau, Hawking and Geroch energies), as operators acting
on the appropriate representation Hilbert space in a particular approach to
quantum gravity (namely, loop quantum gravity).

\subsection{Some remarks on quasi-local angular momentum}
Spinorial techniques provide a natural setting for the
discussion of angular momentum.
This does not only apply to angular momentum, since spinorial and also twistor techniques
define a framework where further quasi-local mass notions can be introduced (e.g. Penrose
mass), and known results can be reformulated in particularly powerful formulations
(e.g. the  discussion of positive mass theorems using the Nester-Witten form).
However, in this article we will not discuss these approaches
and we refer the reader to the relevant sections in Ref. \cite{Szaba04}.
We will focus on certain aspects
of quasi-local expressions for angular momentum of Komar-like type.
As it was shown in section \ref{s:Komar_ang_mom}, choosing a two-sphere ${\cal S}$ in a 3-slice
adapted to the axial symmetry $\phi^\mu$, a 1-form
$L_\mu$ can be found such that the Komar angular momentum
is expressed as
\be
\label{e:Komar-like-angmom}
J(\phi^\mu)=\frac{1}{8\pi}\oint_{\cal S} L_\nu \phi^\nu \sqrt{q}\;d^2x \ \ .
\ee
In particular, in the Komar expression (\ref{e:3+1Komar_ang_mom}) 
we have $L_\mu={q_\mu}^\nu K_{\nu\rho}s^\rho$,
whereas in the spatial infinity expression (\ref{e:glo:angu_mom_def})
this is modified by a term proportional to the trace $K$ of the extrinsic
curvature. The same applies for an angular momentum defined
from the Brown-York charge (\ref{e:BYcharge}) when plugging the expression
 for $j_\mu$ in (\ref{e:BYepsilonj}) into (\ref{e:BYcharge2+1}),
where $\phi^\mu$ does not need to be a symmetry.
The normal fundamental 1-forms $\Omega_\mu$
on ${\cal S}$ (cf. section \ref{s:notation}) provide another avenue to $L_\mu$.
In this section we assume the form (\ref{e:Komar-like-angmom}) for the
angular momentum 
and comment on some approaches to the determination of the 
({\em quasi-symmetry}) axial vector  $\phi^\mu$.

\paragraph{Divergence-free and quasi-Killing axial vectors}
No ambiguity for $\phi^\mu$ is present when an axial symmetry exists 
on ${\cal S}$:
$\phi^\mu$ is taken as the corresponding Killing vector.
In the absence of such a symmetry, we must address two issues. First, expression
(\ref{e:Komar-like-angmom}) depends on the space-like 3-slice in 
which ${\cal S}$ is embedded. This follows from the modification of
the 1-form $L_\mu$ by a total differential under a boost transformation:
$L_\mu\to L_\mu+{}^2\!D_\mu f$ (cf. boost/normalization transformation of $\Omega^{(\ell)}_\mu$
in section \ref{s:notation}).
Angular momentum  can be associated with ${\cal S}$, independently of any
hypersurface $\Sigma$, by demanding the 
axial vector to be divergence-free: ${}^2\!D_\mu \phi^\mu=0$.
Then, the boost-induced modification vanishes under integration.
Second, the physical meaning of $J(\phi^\mu)$ is unclear 
if references to a symmetry notion are completely dropped.
In this sense, different approaches exist aiming at defining appropriate  {\em quasi-Killing}
notions.
We simply mention here some recent works along these lines. In the context of isolated
horizons (see next subsection) a prescription for the determination of a quasi-Killing 
axial vector on black hole horizons has been proposed in  \cite{DreKriSho03}, though 
the divergence-free character is not guaranteed.
Ref. \cite{CooWhi07} presents an approach for finding an approximate Killing vector 
by means of a minimization variational 
prescription that respects the divergence-free character of $\phi^\mu$.
In the context of dynamical or trapping horizons \cite{Haywa94,AshKri02,AshKri04},
a unique divergence-free vector $\phi^\mu$ can be
chosen such that it is preserved by the unique slicing of  
the (space-like) horizon worldtube by marginally outer trapped surfaces \cite{Hay06}.
Also in the context of dynamical horizons, a proposal for $\phi^\mu$
has been made in  \cite{Kor07} relying on a conformal decomposition of the 
metric $q_{\mu\nu}$ on ${\cal S}$. See Ref. \cite{Szaba06} for a discussion
of the divergence-free character of vector fields associated with quasi-local
observables on ${\cal S}$.

Equation (\ref{e:Komar-like-angmom}) only provides 
the expression for the 
component of the angular  momentum vector that is associated with the vector 
$\phi^\mu$.
If we are interested in determining the total angular momentum vector,
a sensible prescription for the other two components
is needed. This is an important practical issue in numerical
simulations (see e.g. \cite{CamLouZlo07}).

\subsection{A study case: quasi-local mass of black hole isolated horizons}
\label{s:IH}
The need of introducing some additional structure has been discussed above in different  
settings (e.g. symmetries for Komar quantities and asymptotic flatness for ADM and Bondi
masses). We illustrate now this issue in a quasi-local context related to
equilibrium black hole horizons. 

\subsubsection{A brief review of isolated horizons}
The isolated horizon framework introduced by Ashtekar and collaborators
\cite{AshKri04} provides a quasi-local setting for characterising black hole horizons
in quasi-equilibrium inside an otherwise dynamical spacetime. It presents
a hierarchical structure with different quasi-equilibrium levels.
The minimal notion of quasi-equilibrium is provided by the so-called
{\em non-expanding horizons} (NEH). Given a Lorentzian manifold, a NEH is a
hypersurface $\mathcal{H}$ such that:
\begin{itemize}
\item[i)] $\mathcal{H}$ is a null hypersurface of topology $S^2 \times \mathbb{R}$ 
that is sliced by marginally (outer) trapped surfaces, i.e. the expansion of the null
congruence associated with the null generator $\ell^\mu$ vanishes on $\mathcal{H}$:
$\theta^{(\ell)}=q^{\mu\nu}\Theta^{(\ell)}_{\mu\nu} = 0$.
\item[ii)] Einstein equation is satisfied on ${\mathcal H}$.
\item[iii)] The vector $-{T^\mu}_\nu\ell^\nu$ is future directed.
\end{itemize}
The geometry of a NEH is characterised by the pair $(q_{\mu\nu}, \hat{\nabla}_\mu)$, where
$q_{\mu\nu}$ is the induced null metric on ${\cal H}$ and $\hat{\nabla}_\mu$ is the
unique connection (not a Levi-Civita one) induced from the ambient spacetime
connection. $\hat{\nabla}_\mu$ characterises the {\em extrinsic geometry} of the NEH. 
A certain combination of components in $\hat{\nabla}_\mu$ can be put together to define
an intrinsic object on $\mathcal{H}$, namely the 1-form $\omega_\mu$ characterised by
\be
\label{e:omega_mu}
\hat{\nabla}_\mu \ell^\nu = \omega_\mu \ell^\nu \ \ .
\ee 
Defining a {\em surface gravity} as  $\kappa_{(\ell)} := \ell^\mu \omega_\mu$,
the acceleration expression for $\ell^\mu$ is given by: 
$\hat{\nabla}_\ell \ell^\mu = \kappa_{(\ell)}\ell^\mu$. On the other hand, 
the projection of $\omega_\mu$ on 
${\cal S}$ recovers the fundamental normal 1-form: $\Omega^{(\ell)}_\mu = {q_\mu}^\rho \omega_\rho$.
The quasi-equilibrium hierarchy is introduced by demanding
the invariance of the null hypersurface  geometry under
the $\ell^\mu$ (evolution) flow in a progressive manner:
\begin{enumerate}
\item A NEH is characterised by the {\em time-invariance} of the
intrinsic geometry $q_{\mu\nu}$: ${\cal L}^{\cal H}_\ell q_{\mu\nu} = 0$.
\item A weakly isolated horizon (WIH) is a NEH, together with 
an equivalence class of null normals $[\ell^\mu]$, for which the 1-form $\omega_\mu$
is time-invariant: ${\cal L}^{\cal H}_\ell \omega_\mu = 0$. This is equivalent
to the (time and angular) constancy of the surface gravity: 
$\hat{\nabla}_\mu \kappa_{(\ell)} = 0$.
\item An isolated horizon (IH) is a WIH on which the whole extrinsic
geometry is time-invariant: $[{\cal L}^{\cal H}_\ell, \hat{\nabla}_\mu]=0$.
\end{enumerate}
The NEH and IH quasi-equilibrium levels represent genuine restrictions
on the geometry of ${\cal H}$ as a hypersurface in the ambient spacetime. On the contrary,
a WIH structure can always be implemented on a NEH by an appropriate choice 
of the null normal $\ell^\mu$  normalization. In this sense, a WIH does
not represent a higher level of quasi-equilibrium than a NEH. However,
from the point of view of the Hamiltonian analysis of spacetimes
with a black hole in quasi-equilibrium as an inner boundary, the WIH notion
proves to be crucial for the correct definition of the phase space symplectic structure
and, more concretely, for the sound formulation
of the quasi-local mass and angular momentum of the horizon.

\subsubsection{An overview of the Hamiltonian analysis of isolated horizons}

\paragraph{Conserved quantities under horizon symmetries}
As in the presentation of ADM quantities in section \ref{s:ADM}, mass and angular 
momentum of isolated horizons
are introduced as conserved quantities under appropriate symmetries
(see \cite{AshFaiKri00,AshBeeLew01} and the 
outline in Appendix C of \cite{GouJar06a} for further details on the following discussion). 
One starts from a symmetry of the horizon structure in the Lorentzian spacetime 
 manifold and then
constructs an associated canonical transformation in the phase or solution
space of the system. The conserved quantity under 
this canonical transformation provides the relevant physical quantity.
In view of the variational problem (see below), a WIH is the relevant horizon structure to be
considered in this
context. A vector field $W^\mu$ preserves the WIH structure
($W^\mu$ is a WIH-symmetry) if
\be
{\cal L}^{\cal H}_W \ell^\mu = \mathrm{const}\cdot \ell^\mu \ \ , \ \
{\cal L}^{\cal H}_W q_{\mu\nu}=0 \ \ ,\ \ {\cal L}^{\cal H}_W \omega_\mu = 0 \ \ .
\ee
WIH-symmetries are of the form $W^\mu = c_W \ell^\mu + b_W S^\mu$,
where $c_W$ and $b_W$ are constants on ${\cal H}$  and $S^\mu$ is 
a Killing vector of any spatial section ${\cal S}$ of  ${\cal H}$.

\paragraph{Variational problem for spacetimes containing WIHs}
In order to set up the Hamiltonian treatment, we need first to define
a well-posed variational problem. Here we are interested
in the variational problem for asymptotically flat spacetimes containing
a WIH. We will furthermore demand this WIH to contain an axial Killing
vector $\phi^\mu$.
The variational problem is then set in the region contained between
two asymptotically Euclidean slices $\Sigma_-$ and $\Sigma_+$, spatial infinity
$i^0$ and the part of ${\cal H}$ between an initial horizon
slice $S_- = {\cal H}\cap \Sigma_-$ and a final one
$S_+ = {\cal H}\cap \Sigma_+$. 
The action, as in Eq. (\ref{e:EH_action}), can be written \cite{AshFaiKri00,AshBeeLew01}
as the sum
of a bulk and a boundary term at spatial infinity, and the variation
of the dynamical fields is set to vanish on the slices $\Sigma_-$ and $\Sigma_+$.
No boundary term associated with the inner
boundary ${\cal H}$  is introduced. The variational problem
is well-posed, in particular the Einstein equation is recovered,
as long as the condition
\be
\label{e:IHwellposed}
\int_{\cal H} \delta \mathbf{\omega} \wedge {}^2\!\mathbf{\epsilon} = 0 \ \ ,
\ee
holds, where $\omega_\mu$ has been introduced in (\ref{e:omega_mu}) and 
${}^2\!{\epsilon}_{\mu\nu}$ is the volume 2-form on sections ${\cal S}$ of ${\cal H}$. 
The crucial ingredient in the well-posedness of the problem 
is precisely the WIH structure.
This is the additional 
structure needed in order to guarantee the vanishing of (\ref{e:IHwellposed}), so that the variational
problem is correctly posed and quasi-local quantities can be defined.

\paragraph{Phase space, canonical transformations and physical quantities}
The phase space is defined by the couple $(\Gamma, \mathbf{J})$
where $\Gamma$ is an infinite-dimensional manifold where each point represents
a solution to the Einstein equation containing a WIH, and  $\mathbf{J}$ is
a symplectic form (a closed 2-form) on $\Gamma$ in terms of which 
the Poisson bracket is defined. In particular, a vector field $X$ on $\Gamma$
generates a canonical transformation if it leaves the symplectic form
invariant: ${\cal L}^\Gamma _ X \mathbf{J}=0$. Using the closedness of $\mathbf{J}$
this is equivalent to the exactness of the 1-form $i_X J$, i.e. to the (local)
existence of a function $H_X$ such that $i_X \mathbf{J} = \delta H_X$ (where
$\delta$ denotes the differential in $\Gamma$). 
In particular, the quantity $H_X$
defined on the phase space is preserved along the flow of $X$. 
In this context, first, the symplectic form can be obtained
from the action by using the {\em conserved symplectic current} method \cite{CrnWit87}
and, second, a vector field $X_W$ on $\Gamma$ can be constructed from a
WIH-symmetry $W^\mu$ on ${\cal H}$ (cf. \cite{AshFaiKri00,AshBeeLew01} for 
details). For the correct definition of a physical parameter associated with a given
WIH-symmetry $W^\mu$, we must assess if the corresponding $X_W$ preserves the canonical form, i.e.
if $i_{X_W}  \mathbf{J}$ is locally exact.
If this is the case, the conserved quantity is simply read from the associated explicit
expression of the Hamiltonian $H_{X_W}$. When this scheme is applied to the axial symmetry
$\phi^\mu$ on ${\cal H}$, the corresponding $X_\phi$ turns out to be automatically
an infinitesimal canonical transformation and the conserved quantity has the form
\be
\label{e:IH_ang_mom}
J_{\mathcal H} :=  X_\phi = \frac{1}{8\pi}\oint_{\Sp_t} \omega_\mu \phi^\mu  \sqrt{q} \; d^2 x 
= \frac{1}{8\pi}\oint_{\Sp_t} \Omega^{(\ell)}_\mu \phi^\mu  \sqrt{q} \; d^2 x\ \ ,
\ee
where $\Sp_t$ is any spatial section of ${\cal H}$. This prescription for 
$J_{\mathcal H}$ exactly coincides with the Komar expression (\ref{e:3+1Komar_ang_mom}).
The mass discussion is more subtle. In this case the WIH-symmetry $t^\mu$ associated with {\em time
evolution} is chosen as an appropriate linear combination of the null normal $\ell^\mu$
and the axial vector $\phi^\mu$. It is then found 
\be
i_{X_t} \mathbf{J} = \delta E_{\mathrm{ADM}} -\left( \frac{\kappa_{(t)}}{8\pi}
\, \delta A_{\mathcal H}
+ \Omega_{(t)} \, 
\delta J_{\mathcal H} \right) \ \ ,
 \ee
where $\kappa_{(t)}$ and $\Omega_{(t)}$ are functions on $\Gamma$ and
$ A_{\mathcal H}$ and $J_{\mathcal H}$ correspond, 
respectively, to the area
of any section of ${\cal H}$ and to the horizon angular momentum in (\ref{e:IH_ang_mom}).
The right-hand-side expression is (locally) exact if 
functions $\kappa_{(t)}$ and $\Omega_{(t)}$ 
depend only on $A_{\mathcal H}$ and $J_{\mathcal H}$, and satisfy:
$\frac{\partial\kappa_{(t)}}{\partial J_{\mathcal H}}=
8\pi\frac{\partial \Omega_{(t)}}{\partial A_{\mathcal H}}$. 
A function
$E^t_{\cal H}$ {\em only} depending on 
$A_{\mathcal H}$ and $J_{\mathcal H}$ then exists, such that we can write
\bea
\label{e:1stBHdyn}
\delta E^t_{\mathcal H} = 
\frac{\kappa_{(t)}(A_{\mathcal H}, J_{\mathcal H})}{8\pi}
\, \delta A_{\mathcal H}
+ \Omega_{(t)}(A_{\mathcal H}, J_{\mathcal H}) \, 
\delta J_{\mathcal H} \ \ .
\eea
To finally determine the quasi-local mass $M_{\cal H}$,
the functional form of $E^t_{\cal H}(A_{\mathcal H}, J_{\mathcal H})$ is normalized to the one 
in the stationary 
Kerr family in $\Gamma$. Note that this is only justified once   
$E^t_{\cal H}$ has been shown to depend {\em only} 
on $A_{\mathcal H}$ and $J_{\mathcal H}$, a non-trivial result.
In sum, for isolated horizons 
$M_{\cal H}(A_{\mathcal H}, J_{\mathcal H}):= M_{\mathrm{Kerr}}(A_{\mathcal H}, J_{\mathcal H})$,
given by the Christodoulou mass expression \cite{Chris70}.
\begin{remark}[Quasi-local first law of black hole dynamics]
Expression (\ref{e:1stBHdyn}) extends the first law of black hole dynamics 
(see section \ref{s:BHthermo})
from the stationary setting to dynamical spacetimes where only the black hole
horizon is in equilibrium.
\end{remark}

\section{Global and quasi-local quantities in black hole physics}
As an application, we briefly comment on some relevant issues concerning
mass and angular momentum in the particular case of black hole spacetimes.

\subsection{Penrose inequality: a {\em claim} for an improved mass positivity result for black holes}
\label{s:Penrose}
In the context of the established gravitational collapse picture, Penrose
\cite{Pen73} proposed an inequality providing an upper bound for the area 
of the spatial sections of black hole event horizons in terms of the square
of the ADM mass. This conjecture followed from
a heuristic chain of arguments including rigorous results (singularity and
black hole uniqueness theorems), together with conjectures such
as weak cosmic censorship and the stationarity of the final state of the evolution of a black hole 
spacetime. A local-in-time version of the Penrose inequality  can be formulated
in terms of data on a Euclidean slice. In this
version Penrose conjecture states that, given an asymptotically Euclidean slice $\Sigma$
containing a black hole under the dominant energy condition, the following inequality 
should be satisfied
\be
A_\mathrm{min} \leq 16 \pi M_\mathrm{ADM}^2 \ \ , 
\ee
where $A_\mathrm{min}$ is the minimal area enclosing the apparent horizon.
In addition, equality is only attained by a slice of 
Schwarzschild spacetime. Though this was originally proposed in an attempt to 
construct counter-examples to the weak cosmic censorship conjecture, 
growing evidence has accumulated supporting its generic validity. Beyond 
spherical symmetry \cite{MalOMu94}, a formal proof 
only exists in the Riemannian case, $K_{\mu\nu}=0$, where
the original derivation \cite{HuiIlm01} (see also \cite{Bra01}) makes use of 
some of the quasi-local expressions 
presented in section \ref{s:quasi-local} (cf. discussion about Geroch and  Hawking energies, that
coincide in this time-symmetric case $K_{\mu\nu}=0$).
The intrinsic geometric relevance of the Penrose inequality 
is reflected in its  alternative name as the {\em isoperimetric inequality
 for black holes} \cite{Gib84}.

Penrose inequality can also
be seen as strengthening the positive ADM mass theorem in section \ref{s:ADM},
for the case of black hole spacetimes: the ADM mass is not only positive
but must be larger than a certain positive-definite quantity.
Though it is tempting to identify this positive quantity 
with some quasi-local mass associated with the black hole, 
e.g. with its irreducible mass $A=: 16\pi M_{\mathrm{irr}}^2$ related 
to the Hawking mass (\ref{e:Hawking}),
a caveat follows from the fact that the relevant minimal 
surface of area $A_\mathrm{min}$ does not necessarily coincide with the 
apparent horizon, as  
examples in \cite{BenDo04} show. In any case, this geometric inequality represents
a bridge between global and quasi-local properties in black hole spacetimes
and has become one of the current geometric and 
physical/conceptual main challenges in General Relativity.

\subsection{Black hole (thermo-)dynamics}
\label{s:BHthermo}
A set of four laws was established in \cite{BarCarHaw73} for stationary black holes.
These black hole laws are analogous in form to the standard thermodynamical laws. Though this
analogy is compelling, the fundamental nature of such relation was only acknowledged under
the light of Hawking's discovery \cite{Hawki75} of the (semiclassical) thermal emission of particles
from the event horizon (Hawking radiation). Given a stationary black hole spacetime with
stationary Killing vector $t^\mu$, black hole rigidity theorems \cite{HawEll73} imply the existence
of a second Killing vector $k^\mu$ that coincides with the null generators $\ell^\mu$ on the horizon.
We can write $k^\mu = t^\mu + \Omega_H \phi^\mu$, where $ \phi^\mu$ is an axial Killing vector and
$\Omega_H$ is a constant referred to as the angular velocity of the horizon (see also \cite{Wald84}).
We can write $k^\nu \nabla_\nu k^\mu = \kappa k^\mu$ on the horizon, which defines 
the surface gravity 
function $\kappa$. The zeroth law of black hole mechanics then states the constancy of the 
surface gravity on the event horizon. The second law, namely Hawking's area 
theorem \cite{Hawki71,Hawki72},
guarantees that the area of the event horizon never decreases, whereas the third law states that
the surface gravity $\kappa$ cannot be reduced to zero in a finite (advanced) time (see \cite{Israe86}
for a precise statement). In the present context, we are particularly interested in the first law,
since it relates the variations of some of the quasi-local and global quantities we have introduced
in the text, in the particular black hole context. First law provides an expression
for the change of the total mass $M$ of the black hole (a well-defined notion since we deal 
with asymptotically flat spacetimes) under a small stationary and 
axisymmetric change in the solution space
\be
\label{e:1stBH}
\delta M = \frac{1}{8\pi} \kappa \delta A + \Omega_H J_H \ \ , \ee
where $A$ is the area of a spatial section of the horizon, and $J_H$
is the Komar angular momentum associated with the axial Killing 
$\phi^\mu$. Equation (\ref{e:1stBH}) relates the variation of a global
quantity $M=M_\mathrm{ADM}$ at spatial infinity on the left-hand-side, to the
variation of quantities locally defined at the horizon, on the
right-hand-side. In particular, we could express the variation of the horizon
area in terms of the variation of the irreducible local mass
$M_{\mathrm{irr}}$, as $\delta A = 32\pi M_{\mathrm{irr}}\delta
M_{\mathrm{irr}}$. Such a formulation actually plays a role in the criterion
for constructing sequences of binary black hole initial data
corresponding to quasi-circular adiabatic inspirals
(cf. \cite{GraGouBon02} and the first law of binary black holes in
\cite{FriUryShi02}).  Derivation of (\ref{e:1stBH}) involves the
notions of ADM mass, as well as the generalization to stationarity of the
Smarr formula for Kerr mass [stating $M = 2 \Omega_H J_H +
\kappa A/(4\pi)$] by using the Komar mass expression. 
Result (\ref{e:1stBHdyn}) in section \ref{s:IH} provides an {\em extension} of
this law to black hole spacetimes non-necessarily stationary, but
containing an isolated horizon for which an unambiguous notion of black
hole mass can be introduced. Quasi-local attempts to extend the first law to 
the fully dynamical case have been explored in the dynamical and trapping
horizon framework \cite{Haywa94,AshKri02,AshKri04,BooFai04}. However, 
the lack of a general
unambiguous notion of quasi-local mass prevents the derivation of a {\it result} 
analogous to (\ref{e:1stBH}) or (\ref{e:1stBHdyn}), i.e. the equality
between the variation of two {\em independent} well-defined quantities. 
In the quasi-local dynamical context, an unambiguous law for the area evolution
can be determined (see e.g. \cite{GouJar06b,BooFai07} and references therein). 
The latter can then be used to {\em define}
a flux of energy through the horizon by comparison with (\ref{e:1stBH}).

\subsection{Black hole extremality: a mass-angular momentum inequality}
Subextremal Kerr black holes are characterised by
presenting  angular momenta bounded by their total masses.
Keeping axisymmetry, it has been recently shown \cite{Dai06a,Chrus07a,Chrus07b}
(see also \cite{DaiOrt09}) that the inequality
\be
\label{e:JMineq}
|J_{\mathrm{K}}|\leq  M_{\mathrm{ADM}}^2 \ \ , 
\ee 
holds also for vacuum, maximal (K=0), axisymmetric Euclidean data. 
Moreover, equality only holds for slices of extremal Kerr.
Inequality (\ref{e:JMineq}) provides a non-trivial relation for black hole spacetimes
between precisely
the two physical quantities we are focusing on in this review.  
It is natural to explore if some analogous inequality holds
when moving  away from axisymmetry and when considering
only the local region around the black hole.
Attempts have been done in this sense, but they all must
face the ambiguities resulting from the absence of 
canonical expressions for quasi-local
masses and angular momenta.
In order to illustrate the caveats to keep in mind
when undertaking this kind of discussion, one can consider the case in which
Komar quantities are used for constructing
a truly quasi-local analogue of expression (\ref{e:JMineq})
for axisymmetric stationary data: initial data have been constructed \cite{AnsPet06} 
where the quotient $|J_{\mathrm{K}}|/M_{\mathrm{K}}^2$ on the black hole horizon 
can become arbitrarily large.
Interestingly, these studies have led to the formulation \cite{AnsPfi07} in axisymmetry
of the related conjecture $8\pi|J_{\mathrm{K}}| \leq A$,
only involving intrinsic quantities on the horizon.
This inequality has been proved to hold in the stationary axisymmetric
case \cite{HenAnsCed08}, as well as in a generalization including the electromagnetic 
field and the associated electric charge
on the horizon \cite{HenAnsCed08b}.

\section{Conclusions}
The problem of characterising the energy-momentum and angular momentum 
associated with the gravitational field in General Relativity has been present
since the birth of the theory and 
controversies have plagued its already long-standing history. Understanding
that no local density of energy-momentum can be identified for the
gravitational field has challenged the validity of the
mass and angular momentum {\em cherished} notions from non-gravitational physics, 
when trying to perform a straightforward extension of these
concepts to the gravitational field in a general relativistic setting.

The study of specific problems suggests
concrete and/or partial solutions.
In this spirit, at low velocities
and weak self-gravities post-Newtonian approaches handle consistent notions
of mass and angular momentum and the same holds in perturbative 
approaches around exact solutions, for which  physical quantities 
can be identified unambiguously. In the same line,
a (quasi-local) notion of the energy carried by a gravitational wave
can be introduced as an {\em average} along the wavelength, proving to 
provide a useful notion in practical applications.
A particular setting of singular conceptual importance is that
of isolated systems in General Relativity, specifically through their characterisation
as asymptotically flat spacetimes. 
The notions of {\em total} ADM and Bondi-Sachs 
energy-momentum provide well-defined quantities that, on the one hand, have
clarified important conceptual issues such as the capability
of gravitational waves to actually carry energy away from a system and, on the other hand,
they also represent 
inestimable tools in practical applications due to their intrinsic/geometric
character. Positivity theorems for the total mass represent 
without any doubt some of the most important and profound
results in General Relativity.
The combination of the success in isolated systems, together
with the absence of a gravitational local energy-momentum density, has led to the 
consideration that
the whole effort for the search of a
{\em local} expression for the gravitational energy represents an ill-posed
or pseudo-problem (see e.g. \cite{MisThoWhe73}). But at the same time, and
motivated by practical needs and/or fundamental physical reasons 
(cf. in this sense \cite{Haag92} for a related discussion on 
{\em quasi-local} issues regarding
observables in Quantum Field Theory), important
efforts have been devoted to the introduction
of quasi-local notions of gravitational energy-momentum 
associated with extended but finite regions of the spacetime. 
In this respect, significant insights into the structure
of the gravitational field have been achieved, 
with applications in diverse conceptual and practical contexts. But
it must be acknowledged, as it is referred in \cite{Szaba04},
that the status of the quasi-local mass studies is in 
a kind of {\em post-modern} situation in which the devoted intensive efforts
have resulted in a plethora of proposals with no obvious definitive
and entirely satisfying candidate. 

A moderate (intermediate) position that avoids
radical skepticism against the quasi-local approach would consist 
in assuming that mass and (in a more restricted sense) angular momentum 
can be unambiguously defined only as global quantities for isolated
systems. But accepting, at the same time, that quasi-local expressions provide 
meaningful and insightful quantities that are inextricably
subject to the need of making systematically 
explicit the specific setting in which they are defined (one can make 
the analogy with the notion of {\em effective mass}
in solid state physics, where  different
masses can be {\em simultaneously} employed for the same particle 
as long as their specific purposes are clearly
stated\footnote{We thank B. Carter for his many comments and insights in this
discussion, and in particular for bringing us to the {\em solid state analogy} for
gravitational masses.}). 
The moral of the whole discussion in this article
is that the formulation
of meaningful global or quasi-local mass and angular momentum notions
in General Relativity {\em always} needs the introduction
of some additional structure in the form of symmetries, quasi-symmetries
or some other background structure. This point must be explicitly
kept in mind whenever employing the so-defined physical quantities, specially
when extrapolating or performing compared analysis.

\medskip

\noindent{\bf Acknowledgements.}
The authors wish to thank the organisers of the Orleans School on Mass 
for their kind invitation and encouragement.

%
%
%
%
%

%
%



\printindex         
\end{document}